\documentclass[11pt,article,bibliography=totoc,bibliography=openstyle, reqno]{article}
\usepackage{amsfonts}
\usepackage{amsthm}
\usepackage{amsmath}
\usepackage{amscd}
\usepackage{amssymb}
\usepackage{subfigure}
\usepackage[T1]{fontenc}            
\usepackage[utf8]{inputenc}       
\usepackage[english]{babel}                  
\usepackage{csquotes}               
\usepackage{t1enc}
\usepackage[mathscr]{eucal}
\usepackage{indentfirst}
\usepackage{graphicx}
\usepackage{booktabs}
\usepackage{pict2e}
\usepackage{color}
\usepackage{epic}
\usepackage{bbm}
\usepackage{enumitem}
\usepackage{tikz}
\usepackage{tkz-graph}
\usetikzlibrary{arrows}
\numberwithin{equation}{section}
\usepackage[margin=2.6cm]{geometry}
\usepackage{epstopdf} 
\usepackage{hyperref}
\usepackage{textcomp} 
\usepackage{autobreak}
\usepackage{subfigure}
\usepackage{float}
\usepackage{dsfont}
\usepackage{mathtools}
\usepackage{pdflscape}
\usepackage{calc}
\usepackage{accents}

\usepackage[backend=bibtex,               
          natbib=true,                 
          block = space,
           bibstyle=authoryear,        
           citestyle=authoryear,        
           babel=other,                 
           maxcitenames=2,              
           maxbibnames=10,              
           url=true,                   
           doi=false,                   
           giveninits=true]{biblatex} 
           
\DeclareNameAlias{sortname}{last-first} 
\DefineBibliographyStrings{german}{andothers={\mbox{et~al.}}} 
\newtheorem{Theorem}{Theorem}[section]
\newtheorem{Lemma}[Theorem]{Lemma}

\newtheorem{Prop}[Theorem]{Proposition}

\newtheorem{Remark}[Theorem]{Remark}
\newtheorem{asum}{Assumption}
\newtheorem{?}[Theorem]{Problem}
\newtheorem{Example}[Theorem]{Example}

\newcommand{\fett}{\boldsymbol}

\numberwithin{equation}{section}

\begin{document}

\begin{center}
\begin{LARGE}
Semi-parametric goodness-of-fit testing for INAR models 
\end{LARGE}\\
\vspace{1cm}
\begin{Large}
Maxime Faymonville\footnote{Department of Statistics,  TU Dortmund University, D-44221 Dortmund, Germany; faymonville@statistik.tu-dortmund.de; corresponding author} \quad Carsten Jentsch\footnote{Department of Statistics,  TU Dortmund University, D-44221 Dortmund, Germany; jentsch@statistik.tu-dortmund.de} \quad Christian H. Weiß\footnote{Department of Mathematics and Statistics,  Helmut-Schmidt-University, D-22008 Hamburg, Germany; weissc@hsu-hh.de}
\end{Large}
\end{center}


\begin{abstract}
Among the various models designed for dependent count data, integer-valued autoregressive (INAR) processes enjoy great popularity.
Typically, statistical inference for INAR models uses asymptotic theory that relies on rather stringent (parametric) assumptions on the innovations such as Poisson or negative binomial distributions.
In this paper, we present a novel semi-parametric goodness-of-fit test tailored for the INAR model class. Relying on the INAR-specific shape of the joint probability generating function, our approach allows for model validation of INAR models without specifying the (family of the) innovation distribution. We derive the limiting null distribution of our proposed test statistic, prove consistency under fixed alternatives and discuss its asymptotic behavior under local alternatives. By manifold Monte Carlo simulations, we illustrate the overall good performance of our testing procedure in terms of power and size properties. In particular, it turns out that the power can be considerably improved by using higher-order test statistics. We conclude the article with the application to three real-world economic data sets. 
\end{abstract}






\section{Introduction}\label{sec:intro} \noindent
Integer-valued autoregressive (INAR) models represent a powerful model class for modeling count time series. They offer a flexible and versatile approach for dealing with autoregressive (AR) time series of non-negative integer values and are the natural analog of the well-known AR model for continuous-valued time series. \citet{duli} introduce the INAR model of order $p$ to follow the recursion 
\begin{align} \label{eq:inarp}
X_t = \alpha_1 \circ X_{t-1} + \ldots + \alpha_p \circ X_{t-p} + \varepsilon_t, \quad t \in \mathbb{Z} = \{\ldots,-1,0,1,\ldots\},
\end{align}
where $\varepsilon_t \overset{\text{i.i.d.}}{\sim} G$, that is, the innovations $(\varepsilon_t,t\in\mathbb{Z})$ are independent and identically distributed and follow a discrete distribution $G$ with range $\mathbb{N}_0=\{0,1,2,\ldots\}$ and probability mass function (pmf) $(G(k), \, k \in \mathbb{N}_0)$. The vector of model coefficients $\boldsymbol{\alpha}=(\alpha_1, \ldots, \alpha_p) \in (0,1)^p$ fulfills $\sum_{j=1}^p \alpha_j < 1$. To ensure the integer-valued modeling of the time series, the model uses the binomial thinning operator \enquote{$\circ$} introduced by \citet{steutel} as 
\begin{align} \label{eq:thinning}
\alpha_j \circ X_{t-j} = \sum\limits_{i=1}^{X_{t-j}} Z_i^{(t,j)}, 
\end{align}
with $(Z_i^{(t,j)}, i \in \mathbb{N}, t \in \mathbb{Z})$, $j \in \{1, \ldots,p\}$ being mutually independent Bernoulli-distributed random variables $Z_i^{(t,j)} \sim \text{Bin}(1, \alpha_j)$ independent of ($\varepsilon_t, \, t \in \mathbb{Z}$). Hence, the thinning operations are independent over time and independent of ($\varepsilon_t, \, t \in \mathbb{Z}$). Additionally, the thinning operation at time $t$ and $\varepsilon_t$ are both independent of $X_s, \, s < t$. These comprehensive independence assumptions are characteristic for the INAR$(p)$ model formulation of \citet{duli}
and they differ from the one of \citet{alosh90}. 
But as only the INAR$(p)$ model of \citet{duli} leads to the traditional Yule--Walker equations for the autocorrelation function (ACF), this model is usually preferred in practice and we focus on this model specification for the remainder of this paper. 
However, for $p=1$, both versions of INAR($p$) models simplify to the INAR(1) model first introduced by \citet{mck} and \citet{alosh}. 

The aforementioned flexibility of INAR models gets lost if one imposes a (parametric) family of distributions for the innovations, which, however, mostly happens in the literature because it simplifies considerably the estimation and the inference for these models. Initially, \citet{alosh} suggested a Poisson distribution for the innovations, which can be considered as the natural analog of the normal distribution in the continuous case. 
However, a Poisson distribution may be too restrictive in applications as it only allows for equidispersion. In the following years, INAR models with several alternative innovation distributions have been considered. For instance, \citet{nbinar1} deal with negative binomial innovations, \citet{jazi_geom} with geometric innovations, \citet{jazi} with a zero-inflated Poisson innovation distribution, and \citet{zoip} with zero-and-one inflated Poisson innovations. But regardless of which innovation distribution we choose, we will always face restrictions and lose some of the flexibility of the INAR($p$) model in \eqref{eq:inarp}. That is why we should not test for restrictive (parametric) null hypotheses $H_0^{\text{para}}$ with \emph{pre-defined} innovation distributions, where 
\begin{align} \label{eq:par_null}
H_0^{\text{para}}: (X_t, \, t \in \mathbb{Z}) \; \text{is INAR($p$) with $G=G_{\lambda}$ for some $\lambda\in\Lambda$}
\end{align}
for some parametric family of innovation distributions $\{G_\lambda, \, \lambda \in \Lambda\}$ with $\Lambda\subset \mathbb{R}^d$ for some (finite) $d$ as it is usually considered in the literature, see e.g.~\citet{meikar}, \citet{hudecova}, \citet{schweer}, \citet{aleksandrov22}, \citet{aleksandrov23} and a bivariate extension in \citet{gof_bivar}. Such parametric assumptions considerably facilitate the estimation and allow for relatively simple testing strategies, but they also make the tests prone to possible model misspecification. Instead, we want to test the (semi-parametric) null hypothesis $H_0^{\text{semi}}$ that the data at hand follow an INAR($p$) model as in \eqref{eq:inarp} with \emph{unspecified} innovation distribution, i.e.
\begin{align} \label{eq:null}
H_0^{\text{semi}}: (X_t, \, t \in \mathbb{Z}) \; \text{is INAR($p$).}
\end{align}
In practice, irrespective of the concrete underlying innovation distribution, it is very helpful to know whether an INAR($p$) process \eqref{eq:inarp} is suitable to adequately capture the dependence structure of the count time series. In this case, a semi-parametric estimation approach can be used. The general relevance of semi-parametric approaches for $\mathbb{Z}$-valued time series was recently demonstrated by \citet{liu}, who consider a similar model setup. In a more general time series setup, \citet{pvqmle} introduce the semi-parametric pseudo-variance quasi-maximum-likelihood estimation which is based on a Gaussian quasi-likelihood function relying on the specification of the pseudo-variance. The latter transfers naturally to time series of bounded counts, which are not considered here. Instead, we use the semi-parametric estimator of \citet{drost}, which does not impose any parametric assumption on the innovation distribution and estimates $G$ non-parametrically. Hence, even without imposing a parametric assumption on the innovation distribution, such (semi-parametric) INAR processes are generally attractive in applications as they are very flexible and still easily interpretable due to their autoregressive nature. 

The paper is organized as follows. We introduce a test statistic for the null hypothesis defined in \eqref{eq:null},  derive its limiting null distribution and derive its asymptotic behavior under fixed and local alternatives in Section \ref{sec:main}. As the limiting null distribution is cumbersome to estimate, we introduce an appropriate bootstrap procedure in Section \ref{sec:bs} to get critical values. Corresponding simulation results are provided in Section \ref{sec:sims}. Three applications to real-world data sets follow in Section \ref{sec:realex}. The paper concludes with Section \ref{sec:concl}, where we summarize the results and give an outlook on possible future research questions. All proofs are deferred to supplementary material \citep{suppl_gof}.

\section{Semi-parametric goodness-of-fit test for INAR models} \label{sec:main} \noindent
Suppose we observe a sample $X_1, \ldots, X_n$ of time-series count data and we want to construct a test statistic for the null hypothesis $H_0^{\text{semi}}$ in \eqref{eq:null}. While \citet{meikar} exclusively consider parametric null hypotheses $H_0^{\text{para}}$ of the form \eqref{eq:par_null} without providing asymptotic theory, we adopt their idea of constructing a suitable $L_2$-type test statistic based on two estimators of the (joint) probability generating function (pgf). The first pgf estimator shall be consistent in general, and the second one \emph{only} under the null $H_0^{\text{semi}}$. In \citet{meikar}, the pgf estimation under their (parametric) null $H_0^{\text{para}}$ facilitates a lot due to their parametric assumption on the innovations, which results in closed form expressions for the pgf, depending on a \emph{finite} number of estimated parameters determining the innovation distribution. In what follows, by contrast, we deal with more general expressions under the semi-parametric setup in \eqref{eq:inarp}. 

\subsection{Joint probability generating function of INAR models} \noindent
As we want to test for INAR-type dependence structure of order $p$, we consider the joint pgf of $p+1$ consecutive random variables $X_t, \ldots, X_{t-p}$, which uniquely determines the full dependence structure of a (stationary) Markov process of order $p$. Then, for $u_0, \ldots, u_p \in [0,1]$, the joint pgf of $X_t, \ldots, X_{t-p}$ is defined as 
\begin{align} \label{pgf_allg}
g_{p}(u_0,  \ldots, u_p) := g_{X_t,  \ldots, X_{t-p}}(u_0,  \ldots, u_p) := E\big(u_0^{X_t}  \cdots u_p^{X_{t-p}}\big).
\end{align}
For the construction of a suitable pgf-based goodness-of-fit test statistic, we exploit the INAR dependence structure of $X_t,\ldots,X_{t-p}$ and derive an explicit representation of the joint pgf $g_p$ for INAR($p$) models.


\begin{Lemma}[Joint pgf of INAR($p$) processes] \label{lemma_pgf_inarp}
For $X_t,  \ldots, X_{t-p}$ following an INAR($p$) process \eqref{eq:inarp}, the pgf $g_p$ defined in \eqref{pgf_allg} can be represented by 
\begin{align} \label{eq:pgfINARp_general}
g_p(u_0 \ldots, u_p)= g_{\varepsilon}(u_0) \cdot E\left(\prod_{j=1}^p \Big\{u_j \big(1+\alpha_j(u_0-1)\big)\Big\}^{X_{t-j}}\right),
\end{align}
where $g_\varepsilon(u_0) =\sum_{k=0}^\infty P(\varepsilon_t=k)\, u_0^k = \sum_{k=0}^\infty G(k)\, u_0^k$.
\end{Lemma}
The proof is contained in Subsection \ref{pr_lemma_pgf_inarp} in the Supplement \citep{suppl_gof}. Taking a closer look at \eqref{eq:pgfINARp_general}, we see that the pgf $g_p$ can be represented as a product of two factors. While the first factor \emph{exclusively} depends on the pmf of the innovation distribution, $(G(k), \, k \in \mathbb{N}_0 )$, the second factor is the joint pgf of (only) $p$ consecutive random variables $X_{t-1},\ldots,X_{t-p}$, whose arguments $u_j (1+\alpha_j(u_0-1))$, $j=1,\ldots,p$ also depend on the autoregressive model coefficients $\alpha_1, \ldots, \alpha_p$.
Note that the representation \eqref{eq:pgfINARp_general} does \emph{not} require any further (parametric) assumptions on $G$.

\subsection{Goodness-of-fit test statistic} \noindent
When we renounce all parametric assumptions on the innovation distribution, the main challenge is the estimation of the INAR model, determined by the model coefficients $\alpha_1, \ldots, \alpha_p$ and the innovation distribution $G$. We cannot resort to (parametric) estimation methods such as moment or (conditional) maximum-likelihood estimation \citep[see e.g.][]{bookweiss} 
that are usually employed to consistently estimate the (low-dimensional) parameter vector determining the innovation distribution. Instead, we use the semi-parametric estimator introduced in \citet{drost}, which maintains the parametric binomial thinning, while simultaneously enabling the \emph{non-parametric} estimation of the innovation distribution. For a small-sample refinement of this semi-parametric estimator using penalization techniques, we refer to \citet{sp_penal}. 

The estimation procedure proposed by \citet{drost} allows for joint estimation of the INAR coefficients and the pmf of the innovation distribution, $(G(k),\, k \in \mathbb{N}_0)$. Given $X_1,\ldots,X_n$, their semi-parametric maximum-likelihood estimator
\begin{align}\label{eq:sp_estimator}
(\boldsymbol{\widehat{\alpha}}_{\text{sp}},\widehat{G}_{\text{sp}}) = (\widehat{\alpha}_{\text{sp},1}, \ldots, \widehat{\alpha}_{\text{sp},p}, \widehat{G}_{\text{sp}}(0), \widehat{G}_{\text{sp}}(1), \ldots),
\end{align} 
which they prove to be consistent and efficient, is defined to maximize the conditional likelihood, i.e.
\begin{align}\label{eq:sp_estimator2}
 (\widehat{\fett{\alpha}}_{\text{sp}}, \widehat{G}_{\text{sp}}) \in \underset{(\fett{\alpha},G) \in [0,1]^p \times \widetilde{\mathcal{G}}}{\text{arg max}} \left( \prod\limits_{t=p+1}^n P^{\fett{\alpha},G}_{(X_{t-1}, \ldots, X_{t-p}), X_t} \right),
\end{align}
where $\widetilde{\mathcal{G}}$ contains all probability measures on $\mathbb{N}_0$ and $P^{\fett{\alpha},G}_{(X_{t-1}, \ldots, X_{t-p}), X_t}$ denotes the transition probabilities 
\begin{align*}
P^{\fett{\alpha},G}_{(x_{t-1}, \ldots, x_{t-p}), x_t} &= \mathbb{P}_{\fett{\alpha}, G} \left( \sum\limits_{j=1}^p \alpha_i \circ X_{t-j}+\varepsilon_t=x_t \mid X_{t-1}=x_{t-1}, \ldots, X_{t-p}=x_{t-p} \right) \\ &= (\text{Bin}(x_{t-1}, \alpha_1) \ast \ldots \ast \text{Bin}(x_{t-p}, \alpha_p) \ast G )\{x_t\}.
\end{align*}
Here, $\mathbb{P}_{\fett{\alpha}, G}$ denotes the underlying probability measure induced by an INAR($p$) process with coefficients $\fett{\alpha}$ and innovation distribution $G$, $\text{Bin}(x_{t-j}, \alpha_j)$ is the binomial distribution with parameters $x_{t-j}$ and $\alpha_j$, $j =1, \ldots, p$ and \enquote{$\ast$} denotes the convolution of distributions. 
The estimator in \eqref{eq:sp_estimator} allows to estimate the pgf in \eqref{eq:pgfINARp_general} under the semi-parametric null $H_0^{\text{semi}}$ in \eqref{eq:null} \emph{without} using any parametric assumption on the innovation distribution. Naturally, we use the plug-in estimator
\begin{align}
\label{eq:pgfINARsp}
\widehat g_{p; H_0}(\mathbf{u}) &:= \widehat g_{\varepsilon}(u_0)\cdot \frac{1}{n-p}\sum_{t=p+1}^{n} \prod_{j=1}^p \Big\{u_j \big(1+\widehat \alpha_{\text{sp},j}(u_0-1)\big)\Big\}^{X_{t-j}},
\end{align}
where $\mathbf{u}=(u_0,u_1,\ldots,u_p)$ and
\begin{align}\label{eq:g_hut_epsilon_u_null}
\widehat g_{\varepsilon}(u_0) &:= \sum_{k=0}^\infty \widehat{P}(\varepsilon_t=k)\, u_0^k = \sum_{k=0}^\infty \widehat G_{\text{sp}}(k)\, u_0^k = \sum_{k=0}^{\max(X_1,\ldots,X_n)} \widehat G_{\text{sp}}(k)\, u_0^k
\end{align}
with the semi-parametric estimators $\widehat{\alpha}_{\text{sp},j}, \, j \in \{1, \ldots, p\}$ and $(\widehat{G}_{\text{sp}}(k), \, k \in \mathbb{N}_0)$ from \eqref{eq:sp_estimator}. The last equality in \eqref{eq:g_hut_epsilon_u_null} holds, because, for fixed $n\in\mathbb{N}$, we have $\widehat{G}_{\text{sp}}(k) = 0 \; \forall k > \max(X_1,\ldots,X_n)$; see \citet{drost} for details. 

While the estimator $\widehat g_{p; H_0}(\mathbf{u})$ explicitly makes use of the INAR structure, the (non-parametric) estimator $\widehat g_{p}(\mathbf{u})$, defined by
\begin{align}
\label{eq:pgfINARp_genest}
\widehat g_{p}(\mathbf{u}):=\frac{1}{n-p} \sum_{t=p+1}^n \prod_{j=0}^p u_j^{X_{t-j}}
\end{align}
is consistent in general, that is, under the null \emph{and} under the alternative. Hence, under the null $H_0^{\text{semi}}$ in \eqref{eq:null} of an underlying INAR($p$) process, both $\widehat g_{p; H_0}(\mathbf{u})$ and $\widehat g_{p}(\mathbf{u})$ estimate the same quantity. This allows to construct the $L_2$-type test statistic $T_n$ defined by
\begin{align} \label{eq:tn_int}
T_n 
=n\int_0^1 \cdots \int_0^1 \Big(\widehat g_{p; H_0}(\mathbf{u})-\widehat g_{p}(\mathbf{u})\Big)^2\, w(\mathbf{u};a)\, d\mathbf{u},
\end{align}
where $d\mathbf{u}:=du_0\cdots du_p$ and $w(\mathbf{u};a) := (a+1)^{p+1} \prod_{j=0}^p u_j^a$ is a weighting function with weighting parameter $a \geq 0$. The weighting function is constructed to integrate to one such that $w$ becomes a probability density function (pdf) on $[0,1]^{p+1}$. Choosing $a=0$ corresponds to no weighting, whereas larger values for $a>0$ (common choices are e.g.~$a=2$ and $a=5$) put more weight close to the right boundaries of the integration intervals $[0,1]$, see \citet{guertler}. 
In what follows, we shall use a more precise notation of $T_n$ than introduced in \eqref{eq:tn_int}. Note that $T_n$ is naturally a function of $X_1,\ldots,X_n$, but also that the definition of $T_n$ relies on (nuisance) parameter estimators $\widehat{\theta}_{\text{sp}}:=(\boldsymbol{\widehat{\alpha}}_{\text{sp}},\widehat{G}_{\text{sp}})$, which are also functions of $X_1,\ldots,X_n$ themselves. This justifies the notations
\begin{align} \label{eq:tn_int_notation}
T_n = T_n(X_1, \ldots, X_n) = T_n(\widehat{\theta}_{\text{sp}}; X_1, \ldots, X_n) = T_n(\widehat{\theta}_{\text{sp}}).
\end{align}
The test statistic $T_n$ in \eqref{eq:tn_int} is of a similar structure as in \citet{meikar}. However, by contrast to their parametric approach, we are using the semi-parametric estimator $\widehat{\theta}_{\text{sp}}=(\boldsymbol{\widehat{\alpha}}_{\text{sp}},\widehat{G}_{\text{sp}})$ from \eqref{eq:sp_estimator}.
We should reject the null hypothesis $H_0^{\text{semi}}$ in \eqref{eq:null} for large values of $T_n$ in \eqref{eq:tn_int}.

\begin{Remark}[Higher-order test statistics] \label{tn_higher_order}
For the construction of test statistics in the spirit of \eqref{eq:tn_int} to test for the null $H_0^{\text{semi}}$ in \eqref{eq:null}, we could also consider pgfs of higher order $s \geq p$, which may be beneficial to better detect Markov chain alternatives of higher order than $p$. That is, for any $s \geq p$, we can define the test statistic 
\begin{align} \label{eq:tns}
T_n^{(s)} =  n\int_0^1 \cdots \int_0^1 \Big(\widehat g_{s; H_0}(u_0,\ldots,u_s)-\widehat g_{s}(u_0,\ldots,u_s)\Big)^2\, w(u_0, \ldots,u_s;a)\, du_0\cdots du_s, \end{align}
where $\widehat g_s$ is defined as in \eqref{eq:pgfINARp_genest} with $p$ replaced by $s$, and
\[\widehat g_{s; H_0}(u_0,\ldots,u_s) :=  \widehat g_{\varepsilon}(u_0)\cdot \frac{1}{n-s}\sum_{t=s+1}^{n} \prod_{j=1}^s \Big\{u_j \big(1+\widehat \alpha_{\text{sp},j}(u_0-1)\big)\Big\}^{X_{t-j}} \]
    with $\widehat \alpha_{\text{sp},j} :=0$ for $j=p+1,\ldots,s$. Note that $T_n^{(p)}=T_n$ holds.
\end{Remark}

While the test statistic $T_n$ as proposed in \eqref{eq:tn_int} requires (numerical) integration, making use of \eqref{eq:pgfINARsp} and \eqref{eq:pgfINARp_genest}, it can also be expressed without any integrals.

\begin{Lemma}[Calculation of $T_n$ without integrals] \label{lemma_tn_a_p}
The test statistic $T_n$ given by \eqref{eq:tn_int} can be equivalently written as 
\begin{align*} 
    T_n &= \frac{n}{(n-p)^2} (a+1)^{p+1} \left( \prod \limits_{j=1}^p \frac{1}{1+X_{t-j}+X_{s-j}+a}\right) \Bigg[ \frac{1}{1+X_t+X_s+a}   \\
    &  + \sum \limits_{k_1,k_2 = 0}^{\max(X_1,\ldots,X_n)} \widehat{G}_{\text{sp}}(k_1)\widehat{G}_{\text{sp}}(k_2)\sum\limits_{i_1=0}^{X_{t-1}+X_{s-1}} \sum\limits_{h_1=0}^{i_1} \ldots \sum\limits_{i_p=0}^{X_{t-p}+X_{s-p}} \sum\limits_{h_p=0}^{i_p} \frac{1}{1+k_1+k_2+a+\sum_{m=1}^p h_m}  \\
    &  \prod\limits_{j=1}^p \binom{X_{t-j}+X_{s-j}}{i_j} \widehat{\alpha}_{\text{sp},j}^{i_j} (-1)^{i_j-h_j} \binom{i_j}{h_j} -2 \sum\limits_{k=0}^{\max(X_1,\ldots,X_n)} \widehat{G}_{\text{sp}}(k)  \\
    &  \sum\limits_{i_1=0}^{X_{t-1}+X_{s-1}} \sum\limits_{h_1=0}^{i_1} \ldots \sum\limits_{i_p=0}^{X_{t-p}+X_{s-p}} \sum\limits_{h_p=0}^{i_p} \frac{1}{1+k+X_s+a+\sum_{m=1}^p h_m} \prod\limits_{j=1}^p \binom{X_{t-j}}{i_j} \widehat{\alpha}_{\text{sp},j}^{i_j} (-1)^{i_j-h_j} \binom{i_j}{h_j}  \Bigg]. \notag
\end{align*}
\end{Lemma}
The proof is contained in Subsection \ref{pr_lemma_tn_a_p} in the Supplement \citep{suppl_gof}.

\subsection{Asymptotic theory} \noindent
To derive the limiting distribution of our test statistic $T_n$, recall that, according to $\eqref{eq:tn_int_notation}$, we are actually confronted with $T_n(\widehat{\theta}_{\text{sp}})$. Before addressing this case in Section \ref{sec:T_n_theta_hat} below, let us consider first the somewhat simpler case of $T_n(\theta_0)$ in Section \ref{sec:T_n_theta}, where the semi-parametric estimator $\widehat{\theta}_{\text{sp}}$ is replaced by some $\theta_0=(\boldsymbol{\alpha}_0,G_0)\in[0,1]^p\times \mathcal{\widetilde G}$.

\subsubsection{Limiting distribution of $T_n(\theta_0)$}\label{sec:T_n_theta} \noindent
Let us consider the test statistic $T_n(\theta_0)=T_n(\theta_0;X_1,\ldots,X_n)$ in more detail. It is defined similarly to $T_n=T_n(\widehat{\theta}_{\text{sp}})$ in \eqref{eq:tn_int}, but with $\widehat g_{p;H_0}$ in \eqref{eq:pgfINARsp} replaced by $g_{p; H_0}$, where
\begin{align*}
g_{p; H_0}(\mathbf{u}) &:= g_{0,\varepsilon}(u_0)\cdot \frac{1}{n-p}\sum_{t=p+1}^{n} \prod_{j=1}^p \Big\{u_j \big(1+\alpha_{0,j}(u_0-1)\big)\Big\}^{X_{t-j}},
\end{align*}
with $g_{0,\varepsilon}(u_0) := 
\sum_{k=0}^\infty G_0(k)\, u_0^k$.
While the original test statistic $T_n(\widehat{\theta}_{\text{sp}})$ introduced in \eqref{eq:tn_int} allows to test for $H_0^{\text{semi}}$ in \eqref{eq:null}, that is, for the \emph{whole} INAR($p$) model class, $T_n(\theta_0)$ is useful to test for null hypotheses of the form
\begin{align} \label{eq:null_theta_fixed}
H_0^{\text{semi}}(\theta_0): (X_t, \, t \in \mathbb{Z}) \; \text{is INAR($p$) with $\theta=\theta_0$}
\end{align}
for $\theta_0=(\boldsymbol{\alpha}_0,G_0)$ with some pre-specified $\boldsymbol{\alpha}_0=(\alpha_{0,1},\ldots,\alpha_{0,p})$ and $G_0=(G_0(k), \, k \in \mathbb{N}_0)$. Note that $H_0^{\text{semi}}(\theta_0)\subset H_0^{\text{semi}}$ for all $\theta_0\in[0,1]^p\times \mathcal{\widetilde G}$.

The following proposition shows that, under the null $H_0^{\text{semi}}(\theta_0)$, the test statistic $T_n(\theta_0)$ can be represented as a degenerate V-statistic, which enables the derivation of its limiting distribution.

\begin{Prop}[$T_n(\theta_0)$ as a V-statistic] \label{Vstat}
Let $\theta_0\in[0,1]^p\times \mathcal{\widetilde G}$. Suppose the null hypothesis $H_0^{\text{semi}}(\theta_0)$ in  \eqref{eq:null_theta_fixed} holds, that is, $X_1, \ldots, X_n$ follow an INAR($p$) process with coefficients $\boldsymbol{\alpha}_0=(\alpha_{0,1},\ldots,\alpha_{0,p})$ and innovation distribution $G_0=(G_0(k), \, k \in \mathbb{N}_0)$. Then, the test statistic $T_n(\theta_0)$ is a degenerate V-statistic. That is, $T_n(\theta_0)$ can be represented as\begin{align*}
T_n(\theta_0) = \frac{n}{(n-p)^2} \sum\limits_{t=p+1}^n \sum\limits_{s=p+1}^n h(Y_t, Y_s; \theta_0),
\end{align*}
where $Y_t=(X_t, \ldots, X_{t-p})$, $Y_s=(X_s, \ldots, X_{s-p})$, $\theta_0=(\boldsymbol{\alpha}_0,G_0)$ and $h: \mathbb{R}^{p+1} \times \mathbb{R}^{p+1} \rightarrow \mathbb{R}$ with
\begin{align} \label{kernel_h}
 h(Y_t,Y_s; \theta_0) =& \int\limits_0^1 \ldots \int \limits_0^1  \left(g_{0,\varepsilon}(u_0) \prod \limits_{j=1}^p (u_j(1+\alpha_{0,j}(u_0-1)))^{X_{t-j}}- \prod \limits_{j=0}^p u_j^{X_{t-j}}\right) \\
 & \times\left(g_{0,\varepsilon}(u_0) \prod \limits_{j=1}^p (u_j(1+\alpha_{0,j}(u_0-1)))^{X_{s-j}}- \prod \limits_{j=0}^p u_j^{X_{s-j}}\right) \notag w(u_0, \ldots,u_p;a) \, du_0 \ldots du_p \notag
 \end{align}
the symmetric and continuous kernel and $E(h(y_t,Y_s;\theta_0))=0$ for all $y_t=(x_t, \ldots, x_{t-p})\in\mathbb{R}^{p+1}$.
\end{Prop}

The proof is contained in Subsection \ref{pr_Vstat} in the Supplement \citep{suppl_gof}.  This finding allows to use the asymptotic results established for degenerate $V$-statistics by \citet{leucht2}, which leads to the following theorem.

\begin{Theorem}[Limiting distribution of $T_n(\theta_0)$ under $H_0^{\text{semi}}(\theta_0)$]\label{limit_distr_prep}
Let $\theta_0\in[0,1]^p\times \mathcal{\widetilde G}$. 
Suppose the null hypothesis $H_0^{\text{semi}}(\theta_0)$ in \eqref{eq:null_theta_fixed} holds, that is, $X_1, \ldots, X_n$ follow an INAR($p$) process with coefficients $\boldsymbol{\alpha}_0=(\alpha_{0,1},\ldots,\alpha_{0,p})$ and innovation distribution $G_0=(G_0(k), \, k \in \mathbb{N}_0)$. Then, for $n \rightarrow \infty$, we have
\begin{align*}
T_n(\theta_0) \overset{d}{\longrightarrow} \sum_{k=1}^\infty \lambda_k Z_k^2,
\end{align*}
where $(Z_k)_k$ is a sequence of independent standard normal random variables and $(\lambda_k)_k$ the sequence of nonzero eigenvalues of the equation
\begin{align}  \label{lambda}
E(h(y,Y_0; \theta_0) \Phi(Y_0)) = \lambda \Phi(y)
\end{align} 
enumerated according their multiplicity, where $(\Phi_k)_k$ are the associated orthonormal eigenfunctions and the kernel $h$ is defined in \eqref{kernel_h}.
\end{Theorem}

The proof is contained in Subsection \ref{pr_limit_distr_prep} in the Supplement \citep{suppl_gof}. Based on the asymptotic results from Theorem \ref{limit_distr_prep}, we can define the (unconditional) test
\begin{align*}
\varphi_{n,\theta_0} := \mathbf{1}_{(q_{1-\gamma},\infty)}(T_n(\theta_0)),    
\end{align*}
where $q_{1-\gamma}$ denotes the $(1-\gamma)$-quantile of the $\chi^2$-type limiting distribution of $\sum_{k=1}^\infty \lambda_k Z_k^2$.




\subsubsection{Limiting distribution of $T_n(\widehat{\theta}_{\text{sp}})$}\label{sec:T_n_theta_hat} \noindent
Now, let us consider again the originally proposed test statistic $T_n := T_n(\widehat{\theta}_{\text{sp}})$ as defined in \eqref{eq:tn_int}. To derive its limiting distribution, we impose the following condition which ensures the consistency of the estimator \eqref{eq:sp_estimator}, see \citet{drost} for details.

\begin{asum} \label{gtildeg}
    We assume that the true parameter $\theta_0=(\boldsymbol{\alpha}_0,G_0) \in A \times \mathcal{G}$, where $\mathcal{G}= \{G \in \widetilde{\mathcal{G}}: 0 < G(0) < 1; E_G\varepsilon_t^{p+4} < \infty\}$ and $A = \{ \boldsymbol{\alpha} \in (0,1)^p: \sum_{j=1}^p \alpha_j < 1 \}$.
\end{asum}

The limiting distribution under the null $H_0^{\text{semi}}$ in \eqref{eq:null} is given in the following theorem. For the proof, we make use of the fact that 
\begin{align}\label{eq:T_n_approx}
T_n(\widehat{\theta}_{\text{sp}})=\widetilde T_n(\theta_0)+o_P(1), 
\end{align}
where $\widetilde T_n$ can be defined as $T_n$ in \eqref{eq:tn_int}, but with kernel $h$ replaced by $\widetilde h$ defined in \eqref{kernel_hhat}, and $\theta_0 = (\boldsymbol{\alpha}_0,G_0) \in A \times \mathcal{G}$. As $\widetilde T_n(\theta_0)$ is also a degenerate V-statistic under $H_0^{\text{semi}}$, we can use again the theory derived in \citet{leucht2} to establish the limiting distribution of $\widetilde T_n(\theta_0)$ and, together with \eqref{eq:T_n_approx}, also of $T_n(\widehat{\theta}_{\text{sp}})$ under the null $H_0^{\text{semi}}$.

\begin{Theorem}[Limiting distribution of $T_n=T_n(\widehat{\theta}_{\text{sp}})$ under $H_0^{\text{semi}}$] \label{limit_distr}
Suppose the null hypothesis $H_0^{\text{semi}}$ in \eqref{eq:null} holds, that is, $X_1, \ldots, X_n$ follow an INAR($p$) process. Let Assumption \ref{gtildeg} be satisfied. Then, for $n \rightarrow \infty$, we have
\begin{align}\label{eq:Tn_limit_null}
T_n \overset{d}{\longrightarrow} \sum_{k=1}^\infty \widetilde{\lambda}_k Z_k^2,
\end{align}
where $(Z_k)_k$ is a sequence of independent standard normal random variables and $(\widetilde{\lambda}_k)_k$ the sequence of nonzero eigenvalues of the equation
\begin{align}  \label{lambda_tilde}
E(\widetilde h(y,Y_0; \theta_0) \widetilde \Phi(Y_0)) = \widetilde \lambda \widetilde \Phi(y)
\end{align} 
enumerated according their multiplicity, where $(\widetilde \Phi_k)_k$ are the associated orthonormal eigenfunctions and the kernel $\widetilde h$ is defined in \eqref{kernel_hhat}.
\end{Theorem}

The proof is contained in Subsection \ref{pr_limit_distr} in the Supplement \citep{suppl_gof}. In the latter, we see that the effect of the estimator $\widehat{\theta}_\text{sp}$ on the limiting distribution is captured by the Fr\'echet derivative
of $E_{\theta_0}(h_1(Y_1,u,\theta))$ with respect to $\theta$ and evaluated at $\theta=\theta_0$ with $h_1$ defined in \eqref{h1}. If this Fr\'echet derivative would be equal to zero, the substitution of $\theta_0$ by $\widehat{\theta}_\text{sp}$ would not affect the limiting distribution at all. But according to \eqref{deriv_alpha} and \eqref{deriv_g}, we see that it is generally not equal to zero and, consequently, the estimator $\widehat{\theta}_\text{sp}$ does affect the limiting distribution of the test statistic.

Now, based on the asymptotic results from Theorem \ref{limit_distr}, we are ready to define the (unconditional) test
\begin{align}\label{test_phi}
\varphi_n := \mathbf{1}_{(\widetilde q_{1-\gamma},\infty)}(T_n),    
\end{align}
where $\widetilde q_{1-\gamma}$ denotes the $(1-\gamma)$-quantile of the $\chi^2$-type limiting distribution of $\sum_{k=1}^\infty \widetilde{\lambda}_k Z_k^2$. As this distribution is not pivotal and cumbersome to estimate, we recommend to use a suitable bootstrap procedure in Section \ref{sec:bs} that was proposed by \citet{jewe}.

\begin{Remark}[Testing parametric null hypotheses $H_0^{\text{para}}$]\label{remark_parametric_testing}
Under the parametric null hypothesis $H_0^{\text{para}}$ in \eqref{eq:par_null}, a test statistic of the form $T_n$ in \eqref{eq:tn_int} can be used. Then,  in \eqref{eq:pgfINARsp}, we have to replace $\widehat g_\epsilon(u_0)$ by $g_{\widehat \lambda,\varepsilon}(u_0) := \sum_{k=0}^\infty G_{\widehat \lambda}(k)\, u_0^k$, where $\widehat \lambda$ is some $\sqrt{n}$-consistent estimator for $\lambda$, $G$ is sufficiently smooth in $\lambda$ and some arbitrary $\sqrt{n}$-consistent estimator $\boldsymbol{\widehat \alpha}$ for $\boldsymbol{\alpha}$ has to be used. Then, its limiting distribution can be derived by similar arguments.
\end{Remark}

\subsection{Power properties} \noindent
In this section, we prove consistency of our proposed test under fixed alternatives in Section \ref{sec_fixed_alternatives} and
discuss its asymptotic behavior under suitable local alternatives in Section \ref{sec_local_alternatives}.

\subsubsection{Power under fixed alternatives}\label{sec_fixed_alternatives} \noindent
Suppose we observe data $X_1,\ldots,X_n$ from some (strictly) stationary count time series process $(X_t,t\in\mathbb{Z})$. Then, we want to test the (semi-parametric) null hypothesis $H_0^{\text{semi}}$ in \eqref{eq:null} against the (natural) alternative
\begin{align} \label{eq:alternative}
H_1^{\text{semi}}: (X_t, \, t \in \mathbb{Z}) \; \text{is \emph{not} an INAR($p$).}
\end{align}
Consequently, $H_1^{\text{semi}}$ consists of all stationary count time series processes that are not INAR processes of some (fixed) order $p$ (with \emph{unspecified} innovation distribution).


Furthermore, as the test statistic $T_n$ in \eqref{eq:tn_int} relies on the semi-parametric estimator $\widehat{\theta}_{\text{sp}}$ of \cite{drost}, who presume an underlying INAR(p) process, we have to make sure that $\widehat{\theta}_{\text{sp}}$ still behaves well if the INAR model is misspecified. That is, in analogy to Theorem 2 in \cite{drost}, we assume that a $\theta_{\text{mis}}=(\boldsymbol{\alpha}_{\text{mis}},G_{\text{mis}})\in(0,1)^p\times \mathcal{G}$ exists such that a weak convergence result
\begin{align}\label{convergence_misspecification}
\sqrt{n}\left(\widehat{\theta}_{\text{sp}}-\theta_{\text{mis}}\right)\rightsquigarrow -\dot \Psi_{\theta_{\text{mis}}}^{\text{mis},-1} \mathbb{S}^{\text{mis}}   
\end{align}
holds, where $\mathbb{S}^{\text{mis}}$ is some tight, Borel measurable Gaussian process obtained as in equation (15) in \cite{drost}, but under the alternative, i.e., when the fitted INAR model is misspecified. Similarly, $\dot \Psi_{\theta_{\text{mis}}}^{\text{mis},-1}$ denotes here the inverse Fr\'echet derivative of the `limiting' estimating equations under the alternative of a misspecifed INAR model, which can be obtained as in equations (8) and (9) in \cite{drost}. 

Hence, for studying the behavior of the test $\varphi_n$ under \emph{fixed} alternatives, let $(X_t,t\in\mathbb{Z})$ be some (strictly) stationary
count time series process 
such that \eqref{convergence_misspecification} holds. Furthermore, we assume that the joint pgf of $X_{t},\ldots,X_{t-p}$, which is denoted by $g_{p}$,
\emph{differs} from $g_{p;H_0}$, which is the joint pgf of the theoretical best INAR($p$) fit to the process $(X_{t},t\in\mathbb{Z})$ under the alternative. More precisely, 
we suppose that 
\begin{align} \label{eq:fixed_alternative_pgf}
g_{p;H_0}(\mathbf{u})-g_{p}(\mathbf{u})=C(\mathbf{u})
\end{align}
holds for some (bounded) function $C:[0,1]^{p+1}\rightarrow \mathbb{R}$ that is non-zero on some (sub)set $S\subseteq [0,1]^{p+1}$ having strictly positive Lebesgue measure $\lambda(S)>0$. Additionally, we assume that the weak convergence
\begin{align}\label{weak_convergence_fixed_alternative}
\big\{\sqrt{n}\left(\widehat g_{p;H_0}(\mathbf{u})-\widehat g_{p}(\mathbf{u})-\left(g_{p;H_0}(\mathbf{u})-g_{p}(\mathbf{u})\right)\right),\mathbf{u}\in U\big\}\Rightarrow \big\{G_{mis}(\mathbf{u}),\mathbf{u}\in U\big\}
\end{align}
holds for some centered Gaussian process $\{G_{mis}(\mathbf{u}),\mathbf{u}\in U\}$ with covariance kernel $K_{mis}(\mathbf{u}_1,\mathbf{u}_2)$, $\mathbf{u}_1,\mathbf{u}_2 \in U$, where $U:=[0,1]^{p+1}$. Then, we have the following result on the asymptotic behavior of $T_n$ under fixed alternatives.

\begin{Theorem}[Consistency of $T_n$ under fixed alternatives] \label{test_cons}
Suppose $(X_t,t\in\mathbb{Z})$ is a strictly stationary 
count time series process
generated under the alternative $H_1^{\text{semi}}$ such that \eqref{convergence_misspecification}, \eqref{eq:fixed_alternative_pgf} and \eqref{weak_convergence_fixed_alternative} hold.
Then, for all $\gamma \in(0,1)$, the test $\varphi_n$ from \eqref{test_phi} is consistent for testing $H_0^{\text{semi}}$ against alternatives $H_1^{\text{semi}}$, that is, $E(\varphi_n)\rightarrow 1$ as $n\rightarrow \infty$.
\end{Theorem}


The proof is contained in Subsection \ref{pr_test_cons} in the Supplement \citep{suppl_gof}. Before we consider the case of local alternatives in the next subsection, Remark \ref{consistency_Markov_p} gives a sufficient Markov condition for \eqref{eq:fixed_alternative_pgf} to hold.

\begin{Remark}[Consistency of $T_n$ for Markov processes of order $p$]\label{consistency_Markov_p}
The class of processes defined by $H_1^{\text{semi}}$ in \eqref{eq:alternative} is too rich to achieve consistency for $T_n$ under general alternatives in $H_1^{\text{semi}}$. For instance, this is
because the stationary distribution of a Markov process of some order $p^\prime>p$ is \emph{not} completely determined by the joint pgf $g_p$ of $X_{t},\ldots,X_{t-p}$. In fact, we would require a test that makes use of the joint pgf $g_{p^\prime}$ of $X_{t},\ldots,X_{t-p^\prime}$, that is, a test of higher order in the sense of Remark \ref{tn_higher_order}. More precisely, by using a higher-order test $T_n^{(s)}$ with $s\geq p^\prime$, consistency is guaranteed in the case of Markovian alternatives of order $p^\prime>p$. 

Hence, when using $T_n=T_n^{(p)}$ for testing the null $H_0^{\text{semi}}$ against $H_1^{\text{semi,Markov$(p)$}}$, where
\begin{align} \label{eq:alternative_Markov_p}
H_1^{\text{semi,Markov$(p)$}}: (X_t, \, t \in \mathbb{Z}) \; \text{is a Markov process of order $p$, but \emph{not} an INAR($p$),}
\end{align}
due to continuity of $g_{p}$ and $g_{p;H_0}$, property \eqref{eq:fixed_alternative_pgf} will generally hold. Consequently, we get consistency of $T_n$ against \emph{all} alternatives in $H_1^{\text{semi,Markov(p)}}$ that also fulfill \eqref{weak_convergence_fixed_alternative}.
\end{Remark} 

In the following example, we illustrate the fixed-alternative setup covered by Theorem \ref{test_cons}.

\begin{Example}[Fixed INAR(2) alternative]\label{Example_fixed_alternatives} 
Suppose we want to use $T_n$ in \eqref{eq:tn_int} (with $p=1$) for testing the null $H_0^{\text{semi}}$ of an INAR(1) process and the data $X_{1},\ldots,X_{n}$ is generated from an INAR(2) process with $\theta=((\alpha_1,\alpha_2), G)$, that is, with coefficients $\alpha_1, \alpha_{2}\in(0,1)$, $\alpha_1+\alpha_2<1$, where $\alpha_{2}\not= 0$, and some innovation distribution $G\in\mathcal{G}$. Then, on the one hand, we have
\begin{align*}
g_{1;H_0}(u_0,u_1)=g_{\varepsilon,H_0}(u_0) \cdot E\left(\Big\{u_1 \big(1+\alpha_{1,H_0}(u_0-1)\big)\Big\}^{X_{t-1}}\right),
\end{align*}
where $g_{\varepsilon,H_0}(u_0) = \sum_{k=0}^\infty G_{H_0}(k)\, u_0^k$ and $\theta_{H_0}=(\alpha_{1,H_0},G_{H_0})$ is the solution of the population analogue of \eqref{eq:sp_estimator2} (i.e.\ the argmax of the expectation of the log-likelihood), when fitting an INAR(1) to $X_{1},\ldots,X_{n}$, when the true DGP $(X_{t},t\in\mathbb{Z})$ is the INAR(2) specified above. On the other hand, using 
$g_1(u_0,u_1)=E(u_0^{X_t}u_1^{X_{t-1}})=E(u_0^{X_t}u_1^{X_{t-1}}1^{X_{t-2}})=g_2(u_0,u_1,1)$,
we have
\begin{align*}
g_{1}(u_0,u_1)
=g_{\varepsilon}(u_0) \cdot E\left(\Big\{u_1 \big(1+\alpha_1(u_0-1)\big)\Big\}^{X_{t-1}}\Big\{\big(1+\alpha_{2}(u_0-1)\big)\Big\}^{X_{t-2}}\right).
\end{align*}
Furthermore, writing $g_{1}(u_0,u_1)=g_{1}(u_0,u_1;\theta)$ and using a Taylor series argument, we have
\begin{align*}
g_{1}(u_0,u_1;\theta) = g_{1}(u_0,u_1;\widetilde \theta_{H_0})+\dot g_{1}(u_0,u_1;\widetilde \theta_{H_0})(\theta-\widetilde \theta_{H_0})+o\left(\frac{1}{\sqrt{n}}\right),
\end{align*}
where $\widetilde \theta_{H_0}=((\alpha_{1,H_0},0),G_{H_0})$. Altogether, due to $g_{1;H_0}(u_0,u_1)=g_{1;H_0}(u_0,u_1;\widetilde\theta_{H_0})=g_{1}(u_0,u_1;\widetilde\theta_{H_0})$, this leads to
\begin{align*}
g_{1;H_0}(u_0,u_1)-g_{1}(u_0,u_1) = -\dot g_{1}(u_0,u_1;\widetilde\theta_{H_0})(\theta-\widetilde\theta_{H_0})+o\left(\frac{1}{\sqrt{n}}\right)
\end{align*}
uniformly in $(u_0,u_1)\in[0,1]^2$. Furthermore, as $\alpha_2\not=0$, we also have $\theta-\widetilde\theta_{H_0}\not=0$ in $[0,1]^2\times \mathcal{G}$ such that 
\begin{align}\label{limit_example_fixed}
n\int_0^1\int_0^1\left(g_{1;H_0}(u_0,u_1)-g_{1}(u_0,u_1)\right)^2w(u_0,u_1;a)du_0du_1\rightarrow +\infty    
\end{align}
as $n\rightarrow \infty$, because $\dot g_{1}(u_0,u_1;\widetilde\theta_{H_0})(\theta-\widetilde \theta_{H_0})$ is non-zero on a set with strictly positive Lebesgue measure such that \eqref{eq:fixed_alternative_pgf} holds.
\end{Example}

\subsubsection{Power under local alternatives}\label{sec_local_alternatives} \noindent
For studying the behavior of the test $\varphi_n$ under \emph{local} alternatives, we have to consider observations from a triangular array of count time series $(X_{n,t}, t=1,\ldots,n,\ n\in\mathbb{N})$. For each fixed $n$, suppose that $X_{n,1},\ldots,X_{n,n}$ are generated from a (strictly) stationary
count time series process under $H_1^{\text{semi}}$ such that the DGP under the alternative depends on $n$ and converges to a DGP under $H_0^{\text{semi}}$ as $n\rightarrow \infty$. We denote the limiting process under $H_0^{\text{semi}}$ by $(X_{0,t},t\in\mathbb{Z})$. For all $n\in\mathbb{N}$, let $g_{p,n}$ with $g_{p,n}(\mathbf{u})=E(u_0^{X_{n,t}}  \cdots u_p^{X_{n,t-p}})$ denote the joint pgf of $X_{n,t},\ldots,X_{n,t-p}$ that \emph{differs} from $g_{p;H_0,n}$, which is the joint pgf of the theoretically best INAR($p$) fit to $(X_{n,t},t\in\mathbb{Z})$. Furthermore, let $\widehat g_{p;H_0,n}(\mathbf{u})$ and $\widehat g_{p,n}(\mathbf{u})$ be the corresponding estimators as defined in \eqref{eq:pgfINARsp} and \eqref{eq:pgfINARp_genest} based on $X_{n,1},\ldots,X_{n,n}$. Note that both quantities are now equipped with an $n$ to match the notation of $g_{p,n}(\mathbf{u})$ and $g_{p;H_0,n}(\mathbf{u})$, but $\widehat g_{p}(\mathbf{u})$ and $\widehat g_{p;H_0}(\mathbf{u})$ defined in \eqref{eq:pgfINARsp} and \eqref{eq:pgfINARp_genest} depend of course already on $n$. Furthermore, 
suppose that $(X_{n,t}, t=1,\ldots,n,\ n\in\mathbb{N})$ is constructed such that
\begin{align} \label{eq:local_alternative_pgf}
g_{p;H_0,n}(\mathbf{u})-g_{p,n}(\mathbf{u})=a_nC(\mathbf{u})+o\left(a_n\right)
\end{align}
holds uniformly over $\mathbf{u}\in U:=[0,1]^{p+1}$ for some $a_n\rightarrow 0$ as $n\rightarrow \infty$ and some (bounded) function $C:[0,1]^{p+1}\rightarrow \mathbb{R}$ that is non-zero on some (sub)set $S\subseteq [0,1]^{p+1}$ having positive Lebesgue measure $\lambda(S)>0$. Additionally, we assume that the weak convergence
\begin{align}\label{weak_convergence}
\big\{\sqrt{n}\left(\widehat g_{p;H_0,n}(\mathbf{u})-\widehat g_{p,n}(\mathbf{u})-\left(g_{p;H_0,n}(\mathbf{u})-g_{p,n}(\mathbf{u})\right)\right),\mathbf{u}\in U\big\}\Rightarrow \big\{G_{loc}(\mathbf{u}),\mathbf{u}\in U\big\}
\end{align}
holds for some centered Gaussian process $\{G_{loc}(\mathbf{u}),\mathbf{u}\in U\}$ with covariance kernel $K_{loc}(\mathbf{u}_1,\mathbf{u}_2)$, $\mathbf{u}_1,\mathbf{u}_2 \in U$. 
Then, we have the following result on the asymptotic behavior of $T_n$ under local alternatives of the form described above. 

\begin{Theorem}[Power of $T_n$ under local alternatives] \label{test_local_power}
Suppose $(X_{n,t}, t=1,\ldots,n,\ n\in\mathbb{N})$ forms a triangular array of count time series and, for each fixed $n$, $X_{n,1},\ldots,X_{n,n}$ is generated from a strictly stationary count time series process
under the alternative $H_1^{\text{semi}}$ such that \eqref{eq:local_alternative_pgf} with $a_n=n^{-1/2}$ and \eqref{weak_convergence} hold. Then, for all $\gamma\in(0,1)$, the test $\varphi_n$ from \eqref{test_phi} fulfills $\lim_{n\rightarrow \infty} E(\varphi_n)=1-F_{\text{loc}}(\widetilde q_{1-\gamma})$, where $F_{\text{loc}}$ denotes the cumulative distribution function of the limiting distribution of $T_n$ under local alternatives, that is, of
\begin{align}\label{eq:limit_local}
\int_0^1 \cdots \int_0^1 \Big(G_{loc}(\mathbf{u})+C(\mathbf{u})\Big)^2 w(\mathbf{u};a) d\mathbf{u}.
\end{align}
If $a_n\rightarrow 0$ such that $\sqrt{n} \, a_n\rightarrow \infty$, the test $\varphi_n$ remains consistent, that is, we have $E(\varphi_n)\rightarrow 1$ as $n\rightarrow \infty$. If $a_n=o(n^{-1/2})$, the test $\varphi_n$ has no asymptotic power, that is, we have $E(\varphi_n)\rightarrow \gamma$ as $n\rightarrow \infty$.
\end{Theorem}
The proof of Theorem \ref{test_local_power} is contained in Subsection \ref{pr_test_local_power} in the Supplement \citep{suppl_gof}. Taking a closer look at the limiting distribution under local alternatives in \eqref{eq:limit_local}, we see that it can be decomposed in three additive terms $A_1$, $A_2$ and $A_3$, where
\begin{align*}
&A_1=\int_0^1 \cdots \int_0^1 G_{loc}^2(\mathbf{u}) w(\mathbf{u};a) d\mathbf{u},   \quad\quad  A_2=2\int_0^1 \cdots \int_0^1 G_{loc}(\mathbf{u})C(\mathbf{u}) w(\mathbf{u};a) d\mathbf{u},    \\
& A_3=\int_0^1 \cdots \int_0^1 C^2(\mathbf{u}) w(\mathbf{u};a) d\mathbf{u}.
\end{align*}
While $A_1$ corresponds to the $\chi^2$-type limiting distribution of $T_n=T_n(\widehat \theta_{\text{sp}})$ derived in \eqref{eq:Tn_limit_null} for the underlying (limiting) process $(X_{0,t},t\in\mathbb{Z})$ under the null $H_0^{\text{semi}}$ and to the first term of $T_n$ discussed in the proof of Theorem \ref{test_cons} for fixed alternatives, the second term $A_2$ has a centered normal distribution with variance determined by the covariance kernel $K_{loc}$ of the Gaussian process $G_{loc}=\{G_{loc}(\mathbf{u}), \mathbf{u}\in[0,1]^{p+1}\}$ and by the function $C(\mathbf{u})$, $\mathbf{u}\in[0,1]^{p+1}$. Finally, the third term $A_3$ is deterministic and strictly positive as the function $C$ is assumed to be non-zero on some set with positive Lebesgue measure. Consequently, together, $A_2+A_3$ has a \emph{non-centered} normal distribution with mean $A_3$ determined by $C$ and variance determined by $G_{loc}$ and $C$. As $A_1$ and $A_2$ are driven by the same Gaussian process $G_{loc}$, $A_1$ and $A_2$ will be typically dependent. Altogether, we see that the limiting distribution of $T_n$ under local alternatives consists of a $\chi^2$-type limiting distribution that is \emph{shifted} by a non-centered normal distribution.


In continuation of Example \ref{Example_fixed_alternatives} dealing with a fixed-alternative setup, we illustrate local alternatives covered 
by Theorem \ref{test_local_power} in the following example.

\begin{Example}[Local INAR(2) alternatives]\label{Example_local_alternatives} 
Suppose we want to use $T_n$ with $p=1$ for testing the null $H_0^{\text{semi}}$ of an INAR(1) process and the data $X_{n,1},\ldots,X_{n,n}$ is generated from an INAR(2) process with coefficients $\alpha_1, \alpha_{2,n}\in(0,1)$, where $\alpha_{2,n}=c/\sqrt{n}$ for some $c\in\mathbb{R}$ such that $\alpha_1,\alpha_{2,n}\in(0,1)$, $\alpha_1+\alpha_{2,n}<1$ for all $n\in\mathbb{N}$ and some innovation distribution $G\in\mathcal{G}$. Then, on the one hand, we have
\begin{align*}
g_{1;H_0,n}(u_0,u_1)=g_{\varepsilon,H_0,n}(u_0) \cdot E\left(\Big\{u_1 \big(1+\alpha_{1,H_0,n}(u_0-1)\big)\Big\}^{X_{t-1}}\right),
\end{align*}
where $g_{\varepsilon,H_0,n}(u_0) = \sum_{k=0}^\infty G_{H_0,n}(k)\, u_0^k$ and $\theta_{H_0,n}=(\alpha_{1,H_0,n},G_{H_0,n})$ is the solution of the population analogue of \eqref{eq:sp_estimator2} (i.e.\ the argmax of the expectation of the log-likelihood), if fitting an INAR(1) to $X_{n,1},\ldots,X_{n,n}$ when the true DGPs of $(X_{n,t},t=1,\ldots,n,\ n\in\mathbb{N})$ are INAR(2) processes. On the other hand, using
\begin{align*}
g_{1,n}(u_0,u_1)=E(u_0^{X_{t,n}}u_1^{X_{t-1,n}})=E(u_0^{X_{t,n}}u_1^{X_{t-1,n}}1^{X_{t-2,n}})=g_{2,n}(u_0,u_1,1),
\end{align*}
we have
\begin{align*}
g_{1,n}(u_0,u_1)
=g_{\varepsilon}(u_0) \cdot E\left(\Big\{u_1 \big(1+\alpha_1(u_0-1)\big)\Big\}^{X_{t-1,n}}\Big\{\big(1+\alpha_{2,n}(u_0-1)\big)\Big\}^{X_{t-2,n}}\right),
\end{align*}
where we used that $g_{\varepsilon,n}(u_0)=g_{\varepsilon}(u_0)$ holds as $G$ does not depend on $n$ by construction. Then, writing $g_{1;H_0,n}(u_0,u_1)=g_{1;H_0,n}(u_0,u_1;\widetilde\theta_{H_0,n})$ where $\widetilde\theta_{H_0,n}=((\alpha_{1,H_0,n},0),G_{H_0,n})\rightarrow \widetilde \theta_0$, and using a Taylor-series argument, we have
\begin{align*}
g_{1;H_0,n}(u_0,u_1;\widetilde\theta_{H_0,n}) = g_{1;H_0,n}(u_0,u_1;\widetilde\theta_0)+\dot g_{1;H_0,n}(u_0,u_1;\widetilde\theta_0)(\widetilde\theta_{H_0,n}-\widetilde\theta_0)+o\left(\frac{1}{\sqrt{n}}\right),
\end{align*}
where $\dot g_{1;H_0,n}(u_0,u_1;\widetilde\theta_0)(\cdot)$ denotes the Fr\'echet derivative of $g_{1;H_0,n}(u_0,u_1;\theta)$ (with respect to $\theta$; see also \eqref{deriv_alpha} and \eqref{deriv_g} in the Supplement \citep{suppl_gof}) evaluated in $\widetilde\theta_0$. Similarly, writing $g_{1,n}(u_0,u_1)=g_{1,n}(u_0,u_1;\theta_n)$, where $\theta_n=((\alpha_1,\alpha_{2,n}),G)\rightarrow ((\alpha_1,0),G)=\widetilde\theta_0$ as $n\rightarrow \infty$, we get
\begin{align*}
g_{1,n}(u_0,u_1;\theta_n) = g_{1,n}(u_0,u_1;\widetilde\theta_0)+\dot g_{1,n}(u_0,u_1;\widetilde \theta_0)(\theta_n-\widetilde\theta_0)+o\left(\frac{1}{\sqrt{n}}\right).
\end{align*}
Altogether, due to $g_{1;H_0,n}(u_0,u_1;\widetilde\theta_0)=g_{1,n}(u_0,u_1;\bar\theta_0)$, this leads to
\begin{align*}
g_{1;H_0,n}(u_0,u_1)-g_{1,n}(u_0,u_1)=\dot g_{1;H_0,n}(u_0,u_1;\widetilde\theta_0)(\widetilde\theta_{H_0,n}-\widetilde\theta_0)-\dot g_{1,n}(u_0,u_1;\widetilde \theta_0)(\theta_n-\widetilde\theta_0)+o\left(\frac{1}{\sqrt{n}}\right).
\end{align*}
Furthermore, we have $\sqrt{n}(\widetilde\theta_{H_0,n}-\widetilde\theta_0)=((0,c),0_{\mathcal{G}})$ and $\sqrt{n}(\theta_n-\bar\theta_0)\rightarrow ((d,0),0_{\mathcal{G}})$ for some $d\in\mathbb{R}$ by construction, where $0_{\mathcal{G}}$ denotes the zero-sequence $0_{\mathcal{G}}=(0,0,0,\ldots)$ $\in 
\mathbb{R}^{\text{dim}(G)}$. Note that the $d$ is in the first entry of the limit, while the $c$ above is in the second entry. Finally, using uniform convergence of \emph{both} 
$\dot g_{1;H_0,n}(u_0,u_1;\widetilde \theta_0)$ and of $\dot g_{1;n}(u_0,u_1;\widetilde \theta_0)$ to $\dot g_{1}(u_0,u_1;\widetilde \theta_0)$, 
we get 
\begin{align}
& n\int_0^1\int_0^1\left(g_{1;H_0,n}(u_0,u_1)-g_{1,n}(u_0,u_1)\right)^2w(u_0,u_1;a)\, du_0\, du_1   \nonumber   \\
\rightarrow& \int_0^1\int_0^1\left(\dot g_{1}(u_0,u_1;\widetilde\theta_0)(((0,c),0_{\mathcal{G}}))-\dot g_{1}(u_0,u_1;\bar \theta_0)(((d,0),0_{\mathcal{G}}))\right)^2w(u_0,u_1;a)\, du_0\, du_1    \label{limit_example_local}
\end{align}
as $n\rightarrow \infty$.
The last right-hand side is strictly positive, because the expressions that is squared and weighted before integration, that is, $\dot g_{1;H_0}(u_0,u_1;\widetilde\theta_0)(((0,c),0_{\mathcal{G}}))-\dot g_{1}(u_0,u_1;\bar \theta_0)(((d,0),0_{\mathcal{G}}))$ is non-zero on a set with strictly positive Lebesgue measure such that \eqref{eq:local_alternative_pgf} holds with $a_n=1/\sqrt{n}$. If $a_n\rightarrow 0$ as a slower rate such that $\sqrt{n} \, a_n\rightarrow \infty$, \eqref{limit_example_fixed} diverges to $+\infty$ as for fixed alternatives in \eqref{limit_example_fixed} such that the corresponding test remains also consistent. If $a_n=o(n^{-1/2})$, $(c,d)$ in \eqref{limit_example_fixed} have to be replaced by $(0,0)$ such that the corresponding test has no asymptotic power as the function to be squared and integrated is exactly zero over $[0,1]^2$.
\end{Example}

\section{Bootstrap inference} \label{sec:bs} \noindent
As we have seen in the previous section and as already stated by \citet{meikar} and beforehand by e.g.~\citet{guertler}, \citet{meiswan}, \citet{leucht}, and \citet{leucht2}, $L_2$-type test statistics as proposed in \eqref{eq:tn_int} do not exhibit a conventional (Gaussian) limiting distribution under the null. Although \citet{drost} derive a CLT for their semi-parametric estimator $(\widehat{\boldsymbol{\alpha}}_{\text{sp}}, \widehat{G}_{\text{sp}})$, this does not lead to a simple limiting distribution of $T_n$, see Theorem \ref{limit_distr}. Therefore, we propose a tailored bootstrap technique to make the testing procedure practicable. On the one hand, the bootstrap has to replicate correctly the binomial thinning operations \eqref{eq:thinning} in the INAR recursion \eqref{eq:inarp} and, on the other hand, we have to use appropriate bootstrap innovations that capture the correct, but unspecified innovation distribution.

Hence, a bootstrap procedure that fulfills these requirements is the semi-parametric INAR bootstrap proposed by \citet{jewe}, which we will outline in the following:
\begin{itemize}
\item[1.)] Fit semi-parametrically an INAR($p$) process \eqref{eq:inarp} using the estimator \eqref{eq:sp_estimator} proposed by \citet{drost} to get estimates $\fett{\alpha}_{\text{sp}}=(\widehat{\alpha}_{\text{sp},1}, \ldots, \widehat{\alpha}_{\text{sp},p})$ and $\widehat{G}_{\text{sp}} = \big(\widehat{G}_{\text{sp}}(k), \, k \in \mathbb{N}_0\big)$ for the INAR coefficients and for the pmf of the innovation distribution, respectively.
\item[2.)] Compute the test statistic $T_n= T_n(\widehat{\theta}_{\text{sp}} ; X_1, \ldots, X_n)$, where $T_n$ is defined in \eqref{eq:tn_int} and $\widehat{\theta}_{\text{sp}}=\widehat{\theta}_{\text{sp}}(X_1,\ldots,X_n)=(\boldsymbol{\widehat \alpha}_{\text{sp}},\widehat G_{\text{sp}})$ is defined in \eqref{eq:sp_estimator}. 
\item[3.)] Generate bootstrap observations $X_1^*, \ldots, X_n^*$ according to \[X_t^* = \widehat{\alpha}_{\text{sp},1} \circ^* X_{t-1}^*+ \ldots + \widehat{\alpha}_{\text{sp},p} \circ^* X_{t-p}^* + \varepsilon_t^*, \] where \enquote{$\circ^*$} denotes (mutually independent) bootstrap binomial thinning operations and $\varepsilon_t^* \overset{\text{i.i.d.}}{\sim} \widehat{G}_{\text{sp}} $ (conditionally on $X_1,\ldots,X_n$).
\item[4.)] Compute the bootstrap test statistic $T_n^*:=T_n(\widehat{\theta}_{\text{sp}}^*; X_1^*, \ldots, X_n^*)$, where $T_n$ is defined in \eqref{eq:tn_int} and $\widehat{\theta}_{\text{sp}}^*=\widehat{\theta}_{\text{sp}}(X_1^*,\ldots,X_n^*)=(\boldsymbol{\widehat \alpha}_{\text{sp}}^*,\widehat G_{\text{sp}}^*)$ is the bootstrap analog of $\widehat{\theta}_{\text{sp}}$ based on $X_1^*, \ldots, X_n^*$.
\item[5.)] Repeat the steps 3.) and 4.) $B$ times, with $B$ sufficiently large, to get bootstrap test statistics $T_n^{*,b}, \, b \in \{1, \ldots, B\}$. 
\item[6.)] Reject the null hypothesis \eqref{eq:null} at significance level $\gamma$ if $T_n=T_n(\widehat{\theta}_{\text{sp}}; X_1,\ldots,X_n)$ exceeds the $(1-\gamma)$-quantile of the empirical distribution of $T_n^{*,b}, \, b \in \{1, \ldots, B\}$.
\end{itemize}

To ensure the stationarity of the bootstrap time series in Step 3 of the above algorithm, we use a burn-in period of $r$ observations, which we will then cut off, i.e., we generate $X_1^*, \ldots, X_{n+r}^*$ and cut the first $r$ values. In the simulation study in Section \ref{sec:sims}, we use $r=100$. To initialize this burn-in period, we use the rounded mean value of the original observations. 

The semi-parametric INAR bootstrap procedure is proved to be (first-order) consistent for the large class of \emph{functions of generalized means} in \cite{jewe} under mild conditions, and it relies on the semi-parametric estimator proposed by \cite{drost}, who proved its (estimation) efficiency. When it comes to formal statements about the efficiency of the bootstrap procedure itself in the sense of higher-order refinements that make typically use of Edgeworth expansions, this would require a lot (more) technical details and is beyond the scope of this paper.

\begin{Remark}[Bootstrap for testing parametric null hypotheses $H_0^{\text{para}}$]
In the situation of a parametric null hypothesis $H_0^{\text{para}}$ \eqref{eq:par_null} discussed in Remark \ref{remark_parametric_testing}, the \emph{parametric} INAR bootstrap of \citet{jewe} finds application. There, the semi-parametric estimators $\widehat{\theta}_{\text{sp}}= (\widehat{\boldsymbol{\alpha}}_{\text{sp}}, \widehat G_{\text{sp}})$ and $\widehat{\theta}_{\text{sp}}^*=(\widehat{\boldsymbol{\alpha}}^*_{\text{sp}}, \widehat G^*_{\text{sp}})$ of Steps 1.) and 4.) are replaced by suitable parametric estimators. 
\end{Remark}

\section{Simulations} \label{sec:sims} \noindent
We investigate the performance of the proposed goodness-of-fit test through a simulation study, where we simulate data from different data generating processes (DGPs) under the null and under the alternative, and where we compute the resulting size and power, respectively. Additionally, we propose a way to better detect violations of the nulls in terms of model order and highlight a big advantage of our semi-parametric test being able to not reject deviations from the null in terms of the INAR \emph{structure}. We mainly focus on the null hypothesis $H_0^{\text{semi}}$ in \eqref{eq:null} with $p=1$ and consider sample sizes $n \in \{100,500\}$. To ensure stationarity of the generated data, we include a prerun of 100 observations which will be omitted afterwards. The significance level $\gamma$ equals $5\%$. We compare our rejection rates with those from the simulation study performed in \citet{meikar}, where the authors considered the parametric null hypothesis \enquote{$H_0: (X_t, \, t \in \mathbb{Z})$ is a Poi-INAR(1) process}, which is of the form $H_0^{\text{para}}$ in \eqref{eq:par_null}. They considered four different DGPs with different parameterizations: INAR(1) with Poi($\lambda$) innovations, INAR(1) with NB($N, \pi$) innovations, Poi-INAR(2) with Poi($\lambda$) innovations, and Poi-INGARCH$(1,1)$. For the latter process, we have $X_t | X_{t-1}, X_{t-2}, \ldots \sim \text{Poi}(M_t)$, where $M_t = \beta_0 + \beta_1 M_{t-1} + \alpha_1 X_{t-1}$. In addition to the DGPs considered in \citet{meikar}, we study further data and test scenarios later on. We consider different weighting parameters $a \in \{0,2,5\}$. For convenience of comparison, we state the rejection rates of \citet[Table~1]{meikar} in italic numbers (where available). They used $a=2$ in their simulations, so we only get direct comparability for this moderate weighting. Since we are concerned with a large computational burden due to the semi-parametric estimation and multiple integration, for conducting bootstrap simulation studies, we use the warp-speed approach with $M=10^4$ Monte--Carlo samples (see \citet{warp} for details). While R \citep{r} is sufficient for moderate sample sizes, also see the \textit{spINAR} package \citep{faymonville2024spinar}, we recommend MATLAB \citep{matlab} for larger $n$. To get an idea of the computing time, see Tables \ref{tab:runtime_poiinar1} and \ref{tab:runtime_poiinar2} in the supplementary material \citep{suppl_gof}, which for different DGPs contain the computing time in seconds for one Monte Carlo sample using the warp speed method and MATLAB.


\subsection{Performance under the null} \noindent
First, we investigate how our test performs under the null $H_0^\text{semi}$ in \eqref{eq:null}. In Table \ref{tab:poiinar1}, we see the rejection rates for a Poi($\lambda$)-INAR(1) DGP with model coefficient $\alpha$ and different weighting parameters $a$. For all weightings, we are rather conservative but we keep the level of $5\%$. \citet{meikar} better exploit the level which could be expected due to their additional (true) information about the innovation distribution. Additionally, to the parameterizations set by \citet{meikar}, we also consider values of $\alpha$ lying closer to the boundaries of its parameter range, i.e., $\alpha \in \{0.1,0.9\}$, see Table \ref{tab:poiinar1_rev} in the Supplement \citep{suppl_gof}. These parameterizations lead to similar results.

Next, we consider the case of an NB($N, \pi$)-INAR(1) DGP. Table \ref{tab:nbinar1} shows that in this case of overdispersion, we are now even closer to the desired size of $5\%$. The (parametric) test of \citet{meikar} will generally reject the INAR model class since this DGP represents a scenario under the alternative for their null. We also test for the null $H_0^\text{semi}$ in \eqref{eq:null} with $p=2$. Table \ref{tab:poiinar2_H0_p2} displays the rejection rates for the different Poi-INAR(2) DGPs. We see that we also keep the level when testing for the null of an INAR process of order 2. Note that \citet{meikar} solely applied their test to the first-order null. 

In all simulation setups, we keep the level of 5\%, but the results are rather conservative. Although \cite{drost}
proved efficiency of their \emph{semi-parametric} INAR model estimator, a large number of (innovations) parameters has to be estimated. In contrast, for the \emph{parametric} INAR model, there will be typically no more than $p+2$ parameters ($p$ INAR coefficients plus one or two parameters for the innovations' mean and variance), which is considerably lower. As these estimators can also leverage the parametric family of innovation distributions, this explains why the semi-parametric test may not hold the level as good as parametric test procedures in finite samples. 

\begin{table}[t]
\centering
\caption{Actual sizes in case of a Poi($\lambda$)-INAR(1) DGP when testing for $H_0^\text{semi}$ in \eqref{eq:null} with $p=1$. Numbers in italic are taken from Table~1 in \citet{meikar} who test for $H_0^\text{para}$ in \eqref{eq:par_null} with $p=1$ and $G_\lambda$=Poi($\lambda$).}
\begin{small}
\begin{tabular}{rr|rr|rr|rr|rr}
\toprule
            &       & \multicolumn{2}{c|}{$a=0$}& \multicolumn{2}{c|}{$a=5$}& \multicolumn{2}{c|}{$a=2$}&\multicolumn{2}{c}{\textit{MK}, $a=2$} \\
 $\lambda$ & $\alpha$ & $n=100$ & $n=500$            & $n=100$ & $n=500$            & $n=100$ & $n=500$            & $n=100$ & $n=500$ \\ 
  \midrule
 1 & 0.3 & 0.036 & 0.033 & 0.037 & 0.040 & 0.036 & 0.035 & \textit{0.049} & \textit{0.055} \\ 
  1 & 0.5 & 0.040 & 0.041 & 0.044 & 0.046 & 0.043 & 0.042 & \textit{0.051} & \textit{0.053} \\ 
  3 & 0.3 & 0.032 & 0.034 & 0.043 & 0.022 & 0.038 & 0.027 & \textit{0.056} & \textit{0.047} \\ 
  3 & 0.5 & 0.032 & 0.032 & 0.037 & 0.029 & 0.039 & 0.034 & \textit{0.055} & \textit{0.059} \\ 
\bottomrule
\end{tabular}
\end{small}
\label{tab:poiinar1}
\end{table}

\begin{table}[t]
\centering
\caption{Actual sizes in case of a NB($N,\pi$)-INAR(1) DGP when testing for $H_0^\text{semi}$ in \eqref{eq:null} with $p=1$. Numbers in italic display the power values taken from Table~1 in \citet{meikar} who test for $H_0^\text{para}$ in \eqref{eq:par_null} with $p=1$ and $G_\lambda$=Poi($\lambda$).}
\begin{small}
\begin{tabular}{rrr|rr|rr|rr|rr}
\toprule
     &        &       & \multicolumn{2}{c|}{$a=0$}& \multicolumn{2}{c|}{$a=5$}& \multicolumn{2}{c|}{$a=2$}& \multicolumn{2}{c}{\textit{MK}, $a=2$} \\
 $N$ & $\pi$ & $\alpha$ & $n=100$ & $n=500$ & $n=100$ & $n=500$ & $n=100$ & $n=500$ & $n=100$ & $n=500$ \\ 
 \midrule
 1 & $1/2$ & 0.5 & 0.048 & 0.050 & 0.050 & 0.053 & 0.052 & 0.052 &\textit{ 0.686 }& \textit{1.000} \\ 
  2 & $2/3$& 0.5 & 0.048 & 0.049 & 0.048 & 0.053 & 0.045 & 0.049 & \textit{0.327} & \textit{0.897 }\\ 
   10 & $10/11$ & 0.5 & 0.047 & 0.046 & 0.049 & 0.046 & 0.046 & 0.045 &\textit{ 0.120} &\textit{ 0.149 }\\ 
\bottomrule
\end{tabular}
\end{small}
\label{tab:nbinar1}
\end{table}

\begin{table}[t]
\centering
\caption{Actual sizes in case of a Poi($\lambda$)-INAR(2) DGP when testing for $H_0^\text{semi}$ in \eqref{eq:null} with $p=2$.}
\begin{small}
\begin{tabular}{rrr|rr|rr|rr}
\toprule
    &        &       & \multicolumn{2}{c|}{$a=0$}& \multicolumn{2}{c|}{$a=5$}& \multicolumn{2}{c}{$a=2$} \\
 $\lambda$ & $\alpha_1$ & $\alpha_2$ & $n=100$ & $n=500$ & $n=100$ & $n=500$ & $n=100$ & $n=500$ \\ 
 \midrule
   1 & 0.3 & 0.1 & 0.037 & 0.040 & 0.036 & 0.024 & 0.035  & 0.024 \\ 
  1 & 0.5 & 0.1 & 0.039 & 0.043 & 0.039 & 0.037 & 0.037  & 0.036 \\ 
 1 & 0.5 & 0.3 & 0.034 & 0.041 & 0.042 & 0.033 & 0.040 & 0.035  \\ 
\bottomrule
\end{tabular}
\end{small}
\label{tab:poiinar2_H0_p2}
\end{table}

\subsection{Performance under the alternative} 
\begin{figure}[t]
\centering
\includegraphics[scale=0.56]{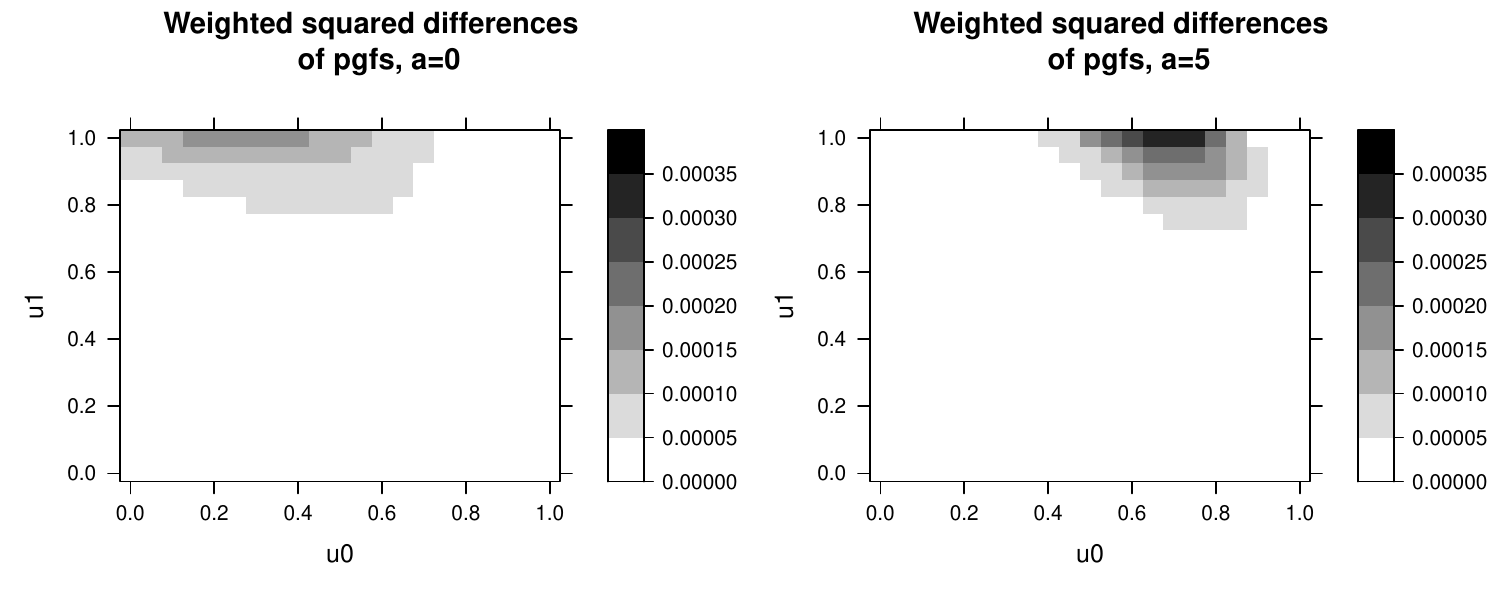} 
\vspace{-0.6cm}
\caption{Heatmaps of the weighted squared difference of the two estimated pgfs for an INGARCH$(1,1)$ DGP (left: $a=0$, right: $a=5$).}
\label{fig:sqdiff_ingarch}
\end{figure}

\noindent
Now, we assess how our test performs when the DGP at hand deviates from the null in terms of model order or model structure. First, we consider the scenario of an INAR(2) with Poi($\lambda$) innovations with the same parameterizations as in Table \ref{tab:poiinar2_H0_p2}. Table \ref{tab:poiinar2} shows that we perform similar to \citet{meikar} though slightly less powerful due to their correctly imposed assumption on the innovations' distribution under the null. With higher weight, however, the power increases, partly surpassing the parametric approach of \citet{meikar}. The same holds in case of an INGARCH$(1,1)$ DGP as demonstrated in Table \ref{tab:ingarch11}. Also in the setup of a Poi-INAR(2) DGP, in addition to the parameterizations considered in \citet{meikar}, we examine settings where $\alpha_1 + \alpha_2$ are close to 0 or 1, see Table \ref{tab:poiinar2_rev_s1} in the Supplement \citep{suppl_gof}. When both $\alpha_1=\alpha_2=0.05$, the test has low power which has been expected since violations of the hypothetical dependence structure are hard to recognize for such a low level of dependence. In particular, the small value of $\alpha_2$ implies that the corresponding INAR(2) process does not differ much from an INAR(1) process. 
In case of $\alpha_1=0.4$ and $\alpha_2=0.5$ however, we achieve even higher power results than for $\alpha_1=0.5$ and $\alpha_2=0.3$ in Table \ref{tab:poiinar2}, explainable through the higher value of $\alpha_2$. Again, the power increases in the weighting parameter $a$. For a better understanding of the weighting concept, consider Figure \ref{fig:sqdiff_ingarch} which contains two heatmaps of $(\widehat g_{1; H_0}(u_0,u_1)-\widehat g_{1}(u_0,u_1))^2w(u_0,u_1;a)$ for a large sample of an INGARCH$(1,1)$ DGP with $\beta_0=1$, $\beta_1=0.1$ and $\alpha_1=0.5$ ($n=5000$). The left heatmap corresponds to $a=0$, i.e.,~no weighting, the right one to $a=5$. We see that the weighting shifts the differences to the upper right corner marking the endpoint of the integration intervals $[0,1]$. In this area, we get darker color, i.e.,~the pgfs differ more. 
Conspicuous at first glance may be the partially poor power results for both the Poi-INAR(2) and the INGARCH$(1,1)$ DGP. In the Poi-INAR(2) case, see Table \ref{tab:poiinar2}, this can be explained by the similarity of an INAR(1) process and an INAR(2) process with small $\alpha_2$. For the INGARCH$(1,1)$ DGPs, \citet{meikar} provide the explanation that such DGPs do not differ much from INAR(1) processes if the ACF at lag 1, i.e., $\rho(1)$, is small. As we can see in Table \ref{tab:ingarch11}, the power distortions exactly occur in such cases of small autocorrelation.  \\

\begin{table}[t]
\centering
\caption{Power in case of a Poi($\lambda$)-INAR(2) DGP when testing for $H_0^\text{semi}$ in \eqref{eq:null} with $p=1$. Numbers in italic are taken from Table~1 in \citet{meikar} who test for $H_0^\text{para}$ in \eqref{eq:par_null} with $p=1$ and $G_\lambda$=Poi($\lambda$).}
\begin{small}
\begin{tabular}{rrr|rr|rr|rr|rr}
\toprule
    &        &       & \multicolumn{2}{c|}{$a=0$}& \multicolumn{2}{c|}{$a=5$}& \multicolumn{2}{c|}{$a=2$}&  \multicolumn{2}{c}{\textit{MK}, $a=2$} \\
 $\lambda$ & $\alpha_1$ & $\alpha_2$ & $n=100$ & $n=500$ & $n=100$ & $n=500$ & $n=100$ & $n=500$ & $n=100$ & $n=500$ \\ 
 \midrule
1 & 0.3 & 0.1 & 0.039 & 0.046 & 0.045 & 0.050 & 0.040 & 0.043 &\textit{ 0.088 }& \textit{0.035 }\\ 
   1 & 0.5 & 0.1 & 0.044 & 0.063 & 0.056 & 0.104 & 0.050 & 0.086 & \textit{0.089 }& \textit{0.127} \\ 
   1 & 0.5 & 0.3 & 0.097 & 0.206 & 0.178 & 0.525 & 0.144 & 0.415 & \textit{0.191} & \textit{0.487} \\ 
\bottomrule
\end{tabular}
\end{small}
\label{tab:poiinar2}
\end{table}

\begin{table}[t]
\centering
\caption{Power in case of an INGARCH$(1,1)$ DGP when testing for $H_0^\text{semi}$ in \eqref{eq:null} with $p=1$. Numbers in italic are taken from Table~1 in \citet{meikar} who test for $H_0^\text{para}$ in \eqref{eq:par_null} with $p=1$ and $G_\lambda$=Poi($\lambda$).}
\begin{footnotesize}
\begin{tabular}{rrrr|rr|rr|rr|rr}
\toprule
     &        &       &  & \multicolumn{2}{c|}{$a=0$}& \multicolumn{2}{c|}{$a=5$}& \multicolumn{2}{c|}{$a=2$}& \multicolumn{2}{c}{\textit{MK}, $a=2$} \\
$\beta_0$ & $\beta_1$ & $\alpha_1$ & $\rho(1)$& $n=100$ & $n=500$ & $n=100$ & $n=500$ & $n=100$ & $n=500$ & $n=100$ & $n=500$\\ 
 \midrule
     0.2 & 0.4 & 0.10 &0.09 & 0.026 & 0.018 & 0.026 & 0.030 & 0.023 & 0.021 & \textit{0.012} &\textit{ 0.050}\\ 
   0.2 & 0.4 & 0.20 &0.21 & 0.031 & 0.087 & 0.054 & 0.131 & 0.044 & 0.117 &\textit{ 0.042 }& \textit{0.108}\\ 
      1.0 & 0.1 & 0.50 &0.52 & 0.068 & 0.205 & 0.159 & 0.645 & 0.113 & 0.501 & \textit{0.152 }& \textit{0.512}\\ 
         0.5 & 0.1 & 0.50 &0.52 & 0.177 & 0.707 & 0.256 & 0.865 & 0.237 & 0.833 &\textit{ 0.248 }&\textit{ 0.724}\\ 
            0.1 & 0.4 & 0.40 &0.53 & 0.264 & 0.829 & 0.255 & 0.816 & 0.264 & 0.825 & \textit{0.296 }& \textit{0.902}\\ 
               0.6 & 0.1 & 0.60 &0.66 & 0.271 & 0.881 & 0.446 & 0.982 & 0.399 & 0.969 &\textit{ 0.376} & \textit{0.988}\\ 
 0.1 & 0.2 & 0.60 &0.70 & 0.497 & 0.985 & 0.443 & 0.973 & 0.468 & 0.980 & \textit{0.534 }& \textit{0.998}\\ 
   0.1 & 0.5 & 0.45 &0.78 & 0.526 & 0.997 & 0.542 & 0.996 & 0.562 & 0.997 & \textit{0.669} & \textit{0.986}\\ 
\bottomrule
\end{tabular}
\end{footnotesize}
\label{tab:ingarch11}
\end{table}

An additional explanation for the generally rather low power values is that when testing the null of an INAR(1) process, we consider the pgf of order 1. However, this does not explain the \emph{entire} dependence structure of an alternative Poi-INAR(2) or INGARCH$(1,1)$ process, since both of them are no first-order Markov chain. To address this, we fit an INAR(1) model to the DGP but now use the second-order test statistic $T_n^{(2)}$ ($s=2$ in \eqref{eq:tns}) by setting $\widehat{\alpha}_{\text{sp},2} :=0$ as suggested in Remark \ref{tn_higher_order}. The size results are presented in Tables \ref{tab:poiinar1_p2} and \ref{tab:nbinar1_p2} in the Supplement \citep{suppl_gof}, where we see that we still keep the level of $5\%$. Tables \ref{tab:poiinar2_p2}, \ref{tab:ingarch11_p2} and \ref{tab:poiinar2_rev_s2} (the latter in the Supplement) contain the resulting power values. For comparison, for the two first DGPs, we again included the power values of \citet{meikar} in italics, which still result from using the first-order ($p=1$) test statistic $T_n$ in \eqref{eq:tn_int}. As anticipated, the power of the second-order test statistics increased, in some settings substantially. Again, we tend to achieve higher power values with higher weighting.

\begin{table}[t]
\centering
\caption{Power in case of a Poi($\lambda$)-INAR(2) DGP when testing for $H_0^\text{semi}$ in \eqref{eq:null} with $p=1$ using test statistic \eqref{eq:tns} with $s=2$. Numbers in italic display the power values taken from Table~1 in \citet{meikar} who test for $H_0^\text{para}$ in \eqref{eq:par_null} with $p=1$ and $G_\lambda$=Poi($\lambda$) not using a higher-order test statistic analogously to \eqref{eq:tns}.}
\begin{small}
\begin{tabular}{rrr|rr|rr|rr|rr}
\toprule
    &        &       & \multicolumn{2}{c|}{$a=0$}& \multicolumn{2}{c|}{$a=5$}& \multicolumn{2}{c|}{$a=2$}&  \multicolumn{2}{c}{\textit{MK}, $a=2$} \\
 $\lambda$ & $\alpha_1$ & $\alpha_2$ & $n=100$ & $n=500$ & $n=100$ & $n=500$ & $n=100$ & $n=500$ & $n=100$ & $n=500$ \\ 
 \midrule
   1 & 0.3 & 0.1 & 0.068 & 0.205 & 0.079 & 0.391 & 0.082 & 0.354&\textit{ 0.088 }& \textit{0.035 } \\ 
  1 & 0.5 & 0.1 & 0.060 & 0.141 & 0.080 & 0.311 & 0.072 & 0.251 & \textit{0.089 }& \textit{0.127}\\ 
 1 & 0.5 & 0.3 & 0.114 & 0.401 & 0.328 & 0.958 & 0.240 & 0.848& \textit{0.191} & \textit{0.487}\\ 
\bottomrule
\end{tabular}
\end{small}
\label{tab:poiinar2_p2}
\end{table}

\begin{table}[t]
\centering
\caption{Power in case of an INGARCH$(1,1)$ DGP when testing for $H_0^\text{semi}$ in \eqref{eq:null} with $p=1$ using test statistic \eqref{eq:tns} with $s=2$. Numbers in italic display the power values taken from Table~1 in \citet{meikar} who test for $H_0^\text{para}$ in \eqref{eq:par_null} with $p=1$ and $G_\lambda$=Poi($\lambda$) not using a higher-order test statistic analogously to \eqref{eq:tns}.}
\begin{footnotesize}
\begin{tabular}{rrrr|rr|rr|rr|rr}
\toprule
     &        &       &  & \multicolumn{2}{c|}{$a=0$}& \multicolumn{2}{c|}{$a=5$}& \multicolumn{2}{c|}{$a=2$}& \multicolumn{2}{c}{\textit{MK}, $a=2$}\\
$\beta_0$ & $\beta_1$ & $\alpha_1$ & $\rho(1)$& $n=100$ & $n=500$ & $n=100$ & $n=500$ & $n=100$ & $n=500$ & $n=100$ & $n=500$ \\ 
 \midrule
     0.2 & 0.4 & 0.10& 0.09 & 0.045 & 0.101 & 0.052 & 0.119 & 0.048 & 0.116 & \textit{0.012} & \textit{0.050 }\\
        0.2 & 0.4 & 0.20 &0.21 & 0.090 & 0.403 & 0.099 & 0.449 & 0.100 & 0.467 &\textit{ 0.042} & \textit{0.108}\\  
           1.0 & 0.1 & 0.50 &0.52 & 0.068 & 0.178 & 0.166 & 0.691 & 0.123 & 0.528 & \textit{0.152} &\textit{ 0.512}\\ 
              0.5 & 0.1 & 0.50 &0.52 & 0.141 & 0.628 & 0.249 & 0.884 & 0.222 & 0.846 & \textit{0.248 }&\textit{ 0.724}\\
                0.1 & 0.4 & 0.40 &0.53 & 0.449 & 0.989 & 0.366 & 0.970 & 0.416 & 0.983 &\textit{ 0.296} & \textit{0.902}\\ 
                   0.6 & 0.1 & 0.60 &0.66 & 0.216 & 0.812 & 0.440 & 0.987 & 0.376 & 0.973 &\textit{ 0.376} &\textit{ 0.988}\\ 
 0.1 & 0.2 & 0.60 &0.70 & 0.566 & 0.996 & 0.489 & 0.989 & 0.534 & 0.994 & \textit{0.534 }& \textit{0.998}\\ 
   0.1 & 0.5 & 0.45 &0.78 & 0.697 & 1.000 & 0.717 & 1.000 & 0.749 & 1.000 & \textit{0.669 }& \textit{0.986}\\ 
\bottomrule
\end{tabular}
\end{footnotesize}
\label{tab:ingarch11_p2}
\end{table}

The parametric testing approaches as in \citet{meikar} allow to test for deviations from the null in terms of model order, which can also be done with our semi-parametric goodness-of-fit test. In addition, due to the flexible and non-restrictive nature of the null hypothesis $H_0^\text{semi}$ in \eqref{eq:null}, we are able to test for deviations from the INAR structure in general.
For this purpose, we consider INARCH(1) and Poi-DAR(1) DGPs with different parameterizations, see \citet{bookweiss}. Both models exhibit an autoregressive structure of order 1, but are distinct from an INAR model. The INARCH(1) process is a special case of the INGARCH$(1,1)$ process, i.e.~$X_t | X_{t-1}, X_{t-2}, \ldots \sim \text{Poi($M_t$), where $M_t = \beta + \alpha X_{t-1}$}$. A Poi-DAR(1) process is characterized by $X_t = a X_{t-1} + b \varepsilon_t$, where $\varepsilon_t \overset{\text{i.i.d.}}{\sim}$ Poi($\lambda$) and $(a,b) \sim \text{Mult}(1, \alpha, 1-\alpha)$. In this latter model class, each observation either chooses the previous observation or the innovation, so the stationary marginal distribution of the innovations equals that of the observations. We choose such values for $\lambda$ and $\beta$ to obtain similar observation means as for the other DGPs. For both DGPs, INARCH(1) and Poi-DAR(1), the power is larger for higher autocorrelation levels (provided that $a > 0$), see Tables \ref{tab:inarch1} and \ref{tab:poidar1}. This is plausible since for high autocorrelation (and small innovation mean), an INAR(1) process tends to produce \enquote{runs}, i.e.,  the same value is realized in consecutive time points. In contrast, the INARCH(1) model shows more erratic behavior. The Poi-DAR(1) process, on the other hand, exhibits even more extreme runs with higher autocorrelation combined with further \enquote{jumps} in-between these runs.

\begin{table}[t]
\centering
\caption{Power in case of an INARCH(1) DGP when testing for $H_0^\text{semi}$ in \eqref{eq:null} with $p=1$.}
\begin{small}
\begin{tabular}{rr|rr|rr|rr}
\toprule
         &       & \multicolumn{2}{c|}{$a=0$}& \multicolumn{2}{c|}{$a=5$}& \multicolumn{2}{c}{$a=2$} \\
 $\beta$ & $\alpha$ & $n=100$ & $n=500$ & $n=100$ & $n=500$ & $n=100$ & $n=500$  \\ 
 \midrule
 1 & 0.30 & 0.032 & 0.044 & 0.048 & 0.129 & 0.039 & 0.094 \\ 
 1 & 0.50 & 0.071 & 0.264 & 0.159 & 0.658 & 0.119 & 0.540 \\ 
 1 & 0.75 & 0.271 & 0.914 & 0.604 & 0.999 & 0.506 & 0.997 \\ 
 3 & 0.30 & 0.035 & 0.030 & 0.046 & 0.022 & 0.038 & 0.025 \\ 
   3 & 0.50 & 0.025 & 0.028 & 0.052 & 0.054 & 0.037 & 0.034 \\ 
3 & 0.75 & 0.074 & 0.184 & 0.185 & 0.634 & 0.133 & 0.460 \\ 
\bottomrule
\end{tabular}
\end{small}
\label{tab:inarch1}
\end{table}

\begin{table}[t]
\centering
\caption{Power in case of a Poi($\lambda$)-DAR DGP when testing for $H_0^\text{semi}$ in \eqref{eq:null} with $p=1$.}
\begin{small}
\begin{tabular}{rr|rr|rr|rr}
\toprule
         &       & \multicolumn{2}{c|}{$a=0$}& \multicolumn{2}{c|}{$a=5$}& \multicolumn{2}{c}{$a=2$} \\
 $\lambda$ & $\alpha$& $n=100$ & $n=500$ & $n=100$ & $n=500$ & $n=100$ & $n=500$  \\ 
 \midrule
 2 & 0.25 & 0.144 & 0.437 & 0.115 & 0.226 & 0.117 & 0.212 \\ 
   2 & 0.50 & 0.447 & 0.992 & 0.535 & 0.995 & 0.538 & 0.994 \\ 
   2 & 0.75 & 0.326 & 0.982 & 0.400 & 0.997 & 0.401 & 0.996 \\ 
   6 & 0.25 & 0.142 & 0.279 & 0.160 & 0.159 & 0.163 & 0.230 \\ 
   6 & 0.50 & 0.117 & 0.371 & 0.234 & 0.707 & 0.190 & 0.643 \\ 
   6 & 0.75 & 0.016 & 0.041 & 0.277 & 0.980 & 0.121 & 0.920 \\ 
\bottomrule
\end{tabular}
\end{small}
\label{tab:poidar1}
\end{table}

\begin{figure}[t]
\centering
\includegraphics[scale=0.65]{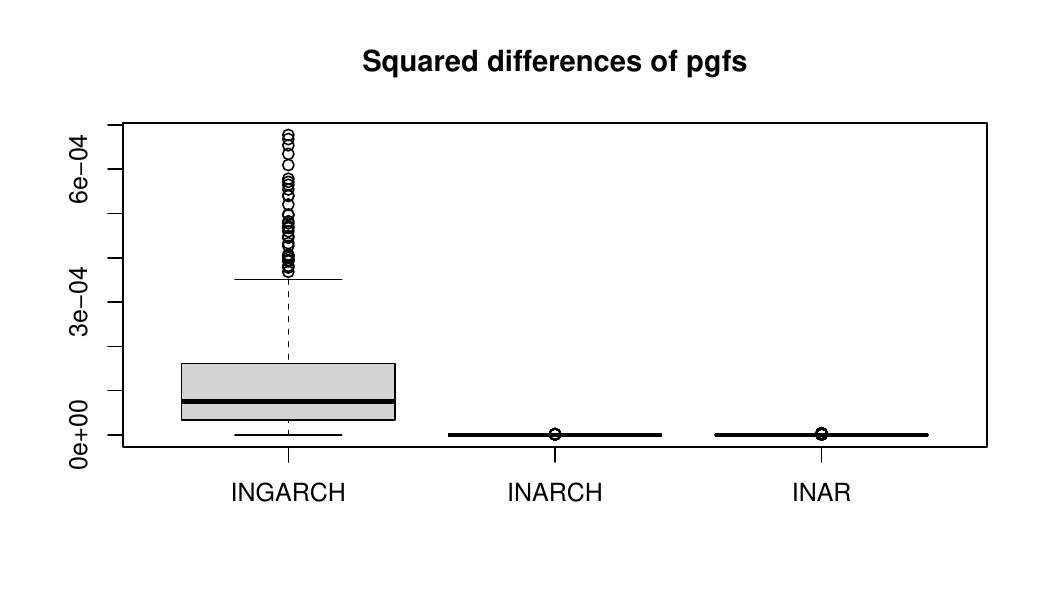} 
\vspace{-0.6cm}
\caption{Boxplots of the squared differences of the two estimated pgfs for the considered INGARCH (left), INARCH (middle) and INAR (right) DGPs.}
\label{fig:boxplots_pgfs}
\end{figure}

While we have high power in most scenarios, we also encounter some parameterizations with low power, e.g., for INARCH(1) processes with low autocorrelation~$\alpha$ and high intercept $\beta$. In general, we seem to lose power for increasing mean of observations (due to additive terms rather than increasing autocorrelation). A higher observation mean leads to a wider range of observations, potentially affecting both the semi-parametric and the non-parametric estimation of the pgf, where the latter is also included in the parametric testing approaches. Besides the challenges related to semi- and non-parametric estimation, the considered DGPs themselves may contribute to low power results. As mentioned at the begin of our paper, without restraining to a certain parametric family of innovations, the INAR model class is very flexible, the unspecified innovation distribution presents a high degree of freedom. Consider for example the INARCH(1) DGP with $\beta=3$ and $\alpha=0.3$ which leads to low power results as displayed in Table \ref{tab:inarch1}. Additionally, consider an INGARCH$(1,1)$ DGP with $\beta_0=0.6$, $\beta_1=0.1$ and $\alpha_1=0.6$ which leads to good power results as displayed in Table \ref{tab:ingarch11}. For both DGPs, we simulated a sample of $n=5000$ observations and computed the integrand of \eqref{eq:tn_int}, $(\widehat g_{1; H_0}(u_0,u_1)-\widehat g_{1}(u_0,u_1))^2$, for $u_0, u_1 \in \{0, 0.05,0.1, \ldots,1\}$. For the sake of comparison, we also considered an INAR(1) DGP with $\lambda=3$ and $\alpha=0.3$. Figure \ref{fig:boxplots_pgfs} shows the boxplots of the resulting values for the three different DGPs. While the two-dimensional pgf of order 1, i.e., $g_2$, of the considered INGARCH$(1,1)$ DGP differs much from the one of a semi-parametrically estimated INAR(1) process, the first-order pgf of the considered INARCH(1) DGP is very close to the latter. Actually, $\widehat g_{1; H_0}(u_0,u_1)$ and $\widehat g_{1}(u_0,u_1)$ do not differ much, the differences are even as small as for the considered INAR(1) DGP. This explains the low power results and underlines the flexibility of the INAR model with unspecified innovation distribution. 

In summary, under the null, it becomes clear that we achieve better results when there is no weighting of the test statistic. However, since a higher weighting of the test statistic leads to substantially higher power values under the alternative and we still keep the level under the null using higher weighting, we recommend using the test with a comparatively high weighting, i.e.~${a=5}$.

\section{Real-world data applications} \label{sec:realex} \noindent
To illustrate the application of our proposed goodness-of-fit test, we apply it on three economic real-world data examples using $B=1000$ bootstrap repetitions. The corresponding MATLAB code is provided in the supplementary material \citep{suppl2_gof}.

\begin{figure}[t]
\centering
\includegraphics[scale=0.5]{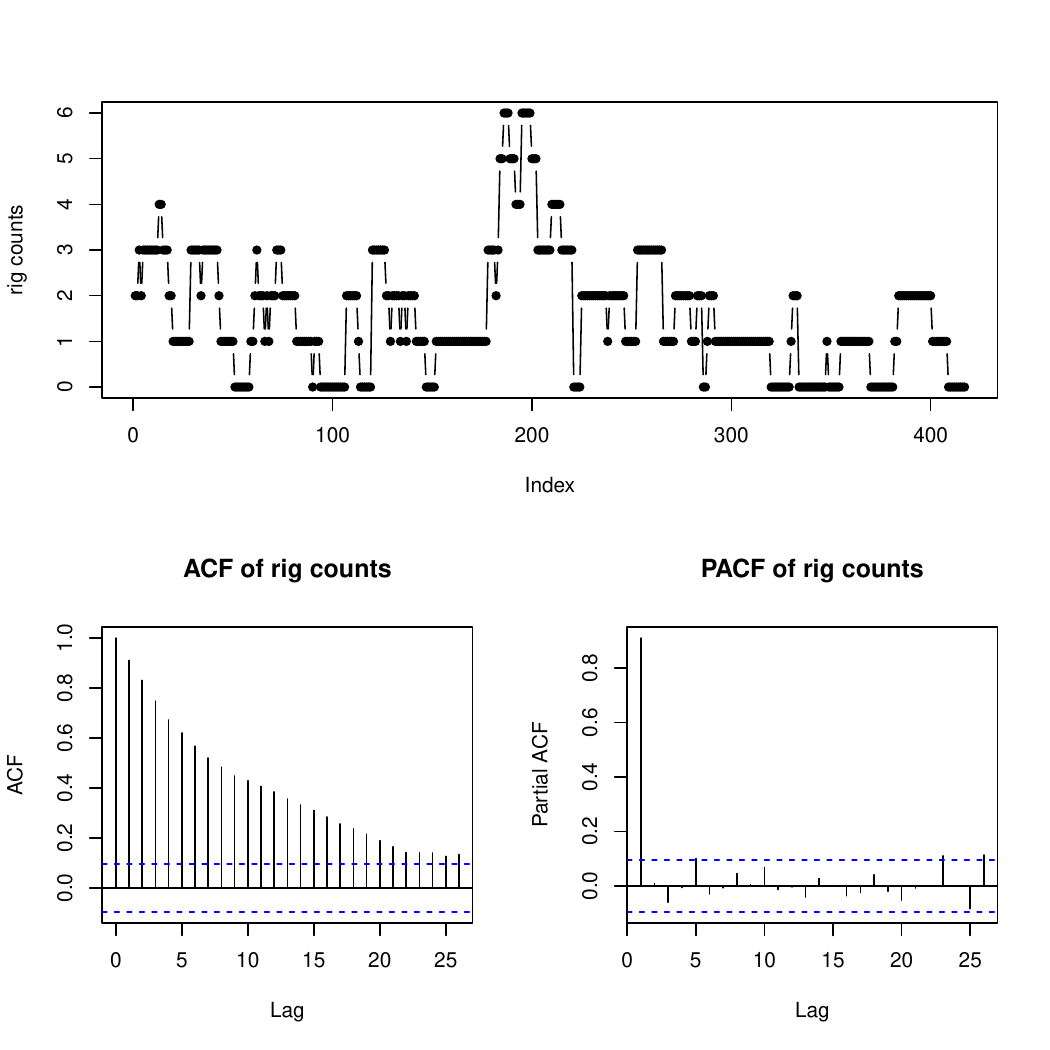} 
\vspace{-0.6cm}
\caption{Plot of rig counts and the corresponding (P)ACF.}
\label{fig:rig}
\end{figure}

The first data set is sourced from Baker Hughes\footnote{phx.corporate-ir.net/phoenix.zhtml?c=79687\&p=irol-rigcountsoverview.} containing weekly counts of active rotary drilling rigs. These counts serve as indicator for the demand of products used in drilling, well completion, oil production, and hydrocarbon processing and have been published since 1944. We specifically focus on the number of drilling rigs in Alaska from 1991 to 1997 ($n=417$). This data set has been addressed before in \citet{bookweiss}. Figure \ref{fig:rig} shows a plot of the time series and the corresponding ACF and PACF. The characteristic INAR runs suggest a high serial dependence with small innovation mean, which are confirmed by the high and slowly decreasing autocorrelation level. Looking at the partial autocorrelation function (PACF), an AR(1)-like model seems to be an appropriate fit. Indeed, when applying our test $T_n$ on the data using $a=5$, we do not reject the null of an INAR(1) process at $5\%$ level. 

\begin{figure}[t]
\centering
\includegraphics[scale=0.5]{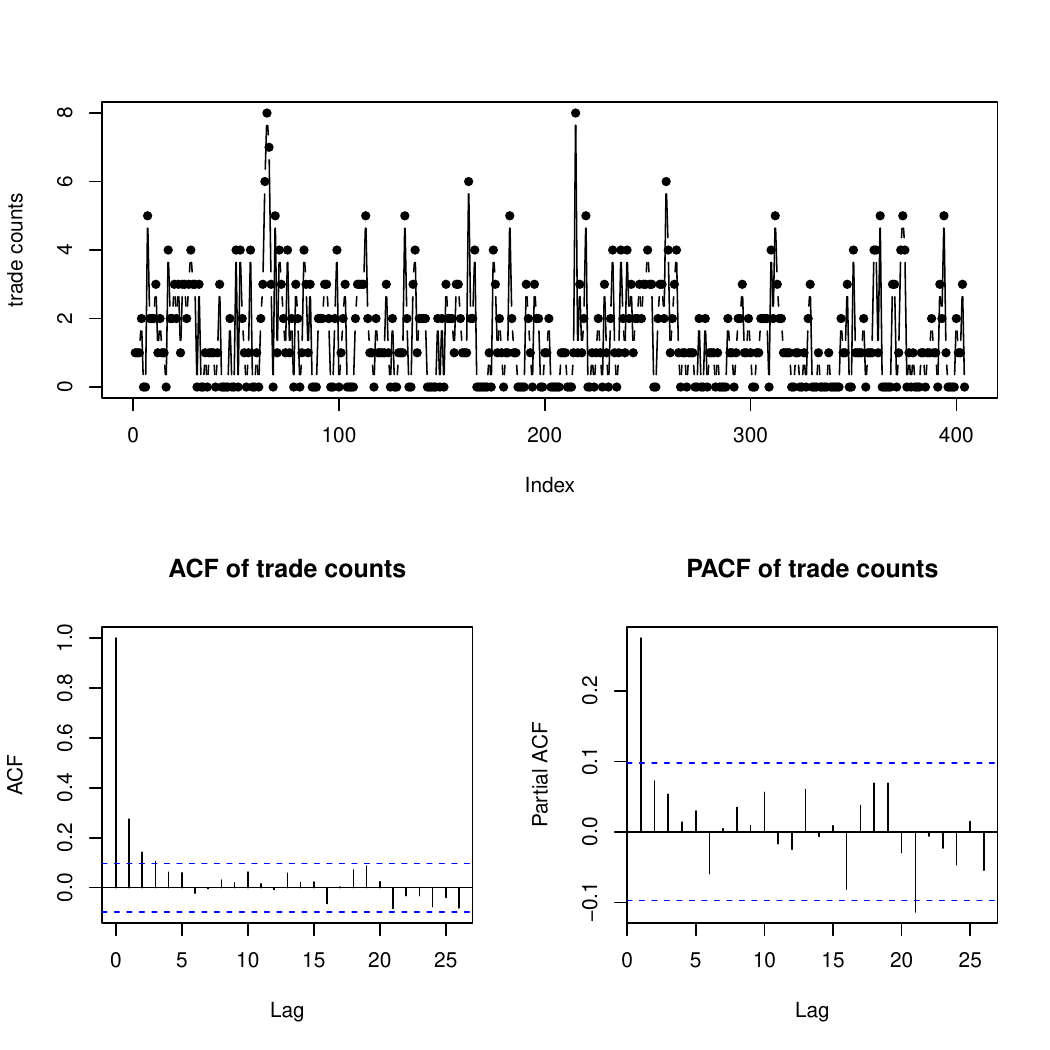} 
\vspace{-0.6cm}
\caption{Plot of daily trade counts and the corresponding (P)ACF.}
\label{fig:tc1_full}
\end{figure}

\begin{figure}[t]
\centering
\includegraphics[scale=0.62]{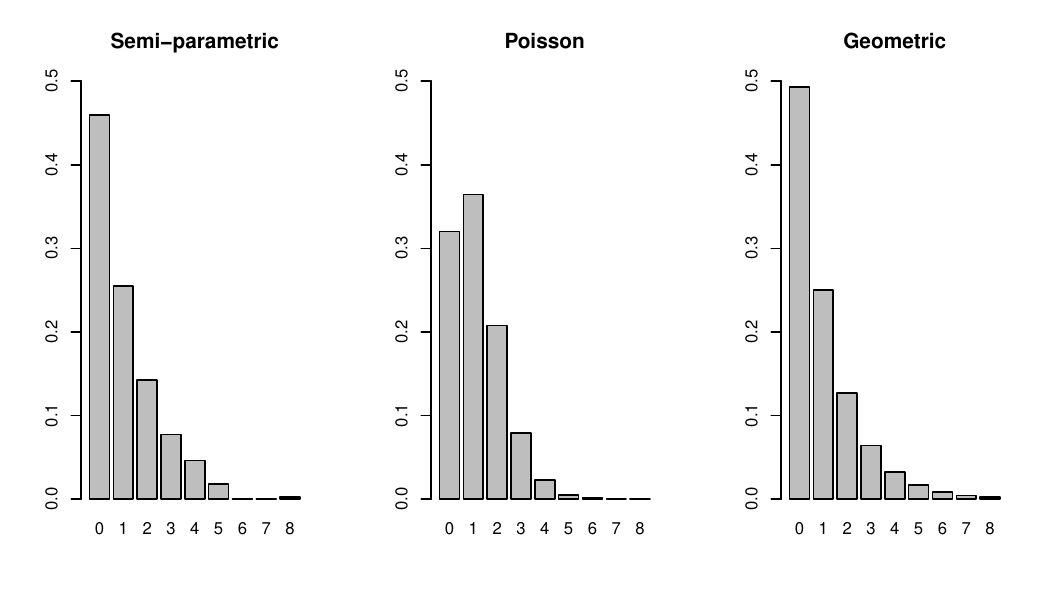} 
\vspace{-0.6cm}
\caption{Plot of the semi-parametrically (left) and parametrically (Poisson in the middle, Geometric on the right) estimated innovation distribution.}
\label{fig:tc1_full_innov}
\end{figure}

The second data set was provided by the Deutsche Börse Group and has been discussed before in \citet{data1}. It contains counts per trading day of transactions of structured products between February 2017 and August 2019 ($n=404$). The data are displayed in Figure \ref{fig:tc1_full}. As in the previous example of rig counts, the ACF and PACF suggest that an INAR(1) model might be an appropriate fit for the data. But applying the test of \citet{meikar}, it rejects the null of a Poi-INAR(1) model at $5\%$ level ($a=2$). Our test, by contrast, also using $a=2$, does not reject the INAR(1) null at $5\%$ level. These different results may be explained by the dispersion of the data. With $\bar{x} \approx 1.47$ and $s^2 \approx 2.23$, the index of dispersion is approximately 1.51 suggesting overdispersed counts. This is additionally stressed out by Figure \ref{fig:tc1_full_innov}. It displays the semi-parametric (left plot) and parametric estimation of the innovation distribution, where for the latter we used a Poisson distribution (in the middle) and a geometric distribution (on the right), respectively.  We see that the parametrically estimated (equidispersed) Poisson distribution differs much from the semi-parametrically estimated innovation distribution whereas the (overdispersed) geometric distribution seems much more appropriate as innovation distribution. Hence, while the parametric test of \citet{meikar} fails to detect the INAR model structure due to their too restrictive equidispersion property of the data under the null, our more flexible test does not reject the INAR structure. 

\begin{figure}[t]
\centering
\includegraphics[scale=0.5]{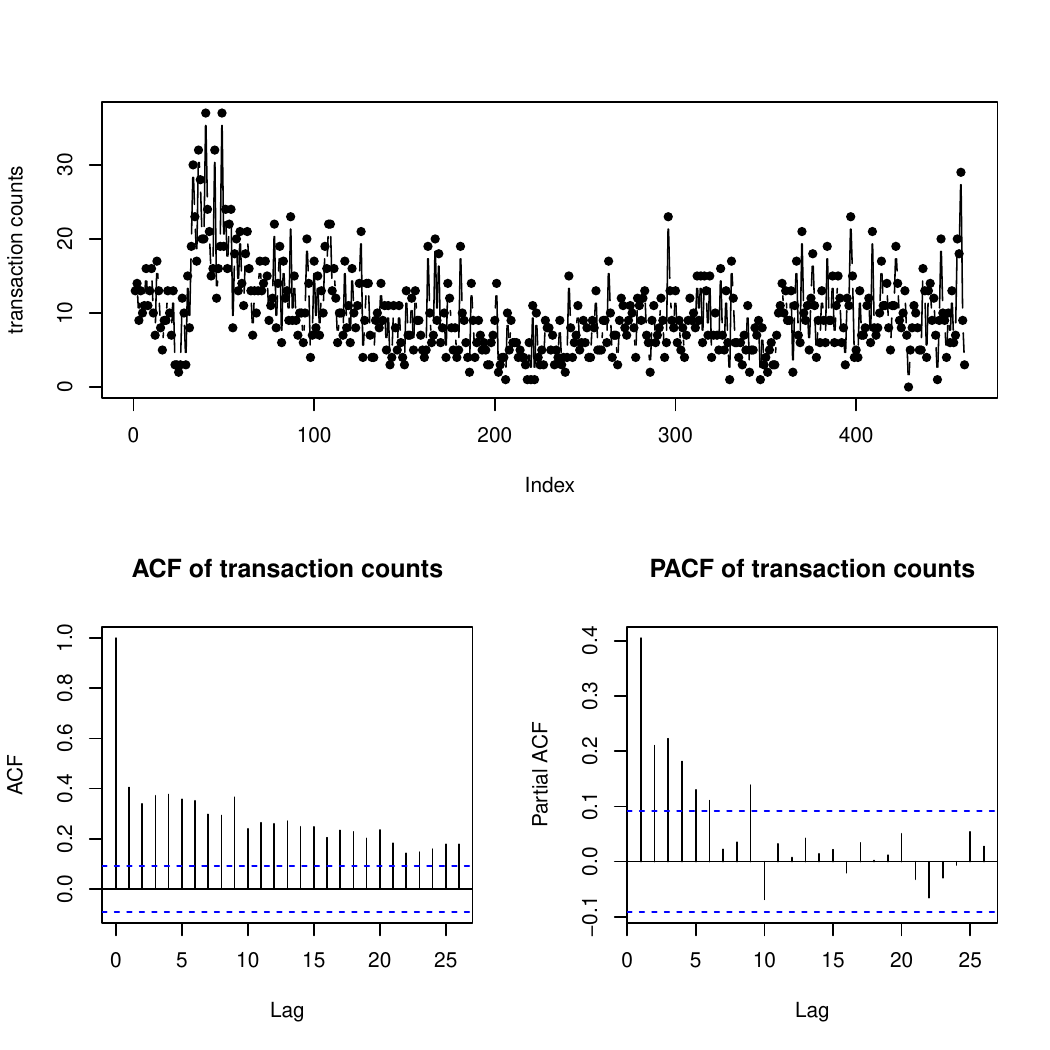} 
\vspace{-0.6cm}
\caption{Plot of transaction counts and the corresponding (P)ACF.}
\label{fig:transa}
\end{figure}

In our third application, we consider a data set first published by \citet{transactions}. It records the number of transaction of the Ericsson B stock per minute between 9:35 and 17:44. Originally, it provides data for the days between July 2 and 22, 2002. \citet{fok09}, \citet{zhu}, \citet{chris_fok}, \citet{davis_liu},\citet{bookweiss} and \citet{su_zhu} exclusively consider the data of July 2 and model the data by an INGARCH$(1,1)$ process. The resulting time series is of length $n=460$. Figure \ref{fig:transa} shows a plot of this time series along with the corresponding (P)ACF. By contrast to the first two examples, this data set exhibits a more erratic structure. The ACF is slowly decaying and the PACF suggests dependencies of higher order than 1. Applying our goodness-of-fit test to the null $H_0^{\text{semi}}$ in \eqref{eq:null} with $p=1$, we are initially not able to reject the null at level $5\%$. 
However, when using the second-order test statistic $T_n^{(2)}$ ($s=2$ in \eqref{eq:tns}), we ultimately reject the null (both with $a=5$). This is plausible since we capture dependencies of higher order by considering the three-dimensional pgf of order 2, i.e., $g_2$, instead of the two-dimensional pgf of order 1, i.e., $g_1$.

\section{Conclusion} \label{sec:concl} \noindent
In existing literature, goodness-of-fit tests for the INAR model class are restricted to specific parametric families of innovation distributions. In this paper, we introduced a novel goodness-of-fit test for INAR($p$) processes that does not rely on parametric assumptions about the nature of the innovations. We derived the null limiting distribution of our $L_2$-type test statistic based on weighted integrals using probability generating functions. Additionally, we proved its consistency under fixed alternatives, discussed the asymptotic behavior under local alternatives, and specified a bootstrap procedure required to circumvent the complex limiting distribution. In an extensive simulation study, we compared our proposed procedure with the parametric competitor of \citet{meikar}. Overall, we got similar results, but we were able to increase the power against Markov chain alternatives of higher order by using also higher-order test statistics.
Moreover, unlike the parametric approaches of \citet{meikar}, we are able to test for general deviations from the INAR structure and got good power results for the considered DGPs. We noticed that when testing with higher weight parameter $a$, the test exhibited higher power.
To conclude the paper, we applied our method to three real data examples from economics.


\section*{Acknowledgments} \noindent
The authors thank the editor and the two referees for their useful comments on an earlier draft of this article. 
The authors gratefully acknowledge the computing time provided on the Linux HPC cluster at TU Dortmund University (LiDO3), partially funded in the course of the Large-Scale Equipment Initiative by the German Research Foundation (DFG) as project 271512359.

\section*{Funding} \noindent
This research was funded by the Deutsche Forschungsgemeinschaft (DFG, German Research Foundation) - Projeknummer 437270842.

\section*{Supplement I} \noindent
We provide additional tables and all the proofs of this paper.

\section*{Supplement II} \noindent
We provide the MATLAB code for the real-world data applications.



\newpage

\section*{References}
\printbibliography[title=References, heading = none]
\newpage


\newpage

\begin{appendix}

\section{Additional tables}

\begin{table}[h]
\centering
\caption{Computing time (in seconds) in case of a Poi-INAR(1) DGP, testing for $H_0^{\text{semi}}$ in (1.4) with $p=1$.}
\begin{small}
\begin{tabular}{rr|rr|rr|rr}
\toprule
            &       & \multicolumn{2}{c|}{$a=0$}& \multicolumn{2}{c|}{$a=5$}& \multicolumn{2}{c}{$a=2$} \\
 $\lambda$ & $\alpha$ & $n=100$ & $n=500$            & $n=100$ & $n=500$            & $n=100$ & $n=500$            \\ 
  \midrule
 1 & 0.3 & 0.109 & 0.176 & 0.071 & 0.168 & 0.059 & 0.184  \\ 
  1 & 0.5 & 0.060 & 0.214 & 0.050 & 0.166 & 0.058 & 0.164  \\ 
  3 & 0.3 & 0.103 & 0.254 & 0.091 & 0.256 & 0.076 & 0.275  \\ 
  3 & 0.5 & 0.083 & 0.281 & 0.100 & 0.230 & 0.150 & 0.312  \\ 
\bottomrule
\end{tabular}
\end{small}
\label{tab:runtime_poiinar1}
\end{table}

\begin{table}[h]
\centering
\caption{Computing time (in seconds) in case of a Poi-INAR(2) DGP, testing for $H_0^{\text{semi}}$ in (1.4) with $p=2$.}
\begin{small}
\begin{tabular}{rrr|rr|rr|rr}
\toprule
    &        &       & \multicolumn{2}{c|}{$a=0$}& \multicolumn{2}{c|}{$a=5$}& \multicolumn{2}{c}{$a=2$} \\
 $\lambda$ & $\alpha_1$ & $\alpha_2$ & $n=100$ & $n=500$ & $n=100$ & $n=500$ & $n=100$ & $n=500$ \\ 
 \midrule
   1 & 0.3 & 0.1 & 5.450 & 11.910 & 2.970 & 6.344 & 4.424  & 11.102 \\ 
  1 & 0.5 & 0.1 & 5.611 & 18.004 & 3.630 & 9.614 & 4.545  & 16.480 \\ 
 1 & 0.5 & 0.3 & 8.123 & 32.553 & 4.396 & 17.867 & 6.796 & 19.555  \\ 
\bottomrule
\end{tabular}
\end{small}
\label{tab:runtime_poiinar2}
\end{table}

\begin{table}[h]
\centering
\caption{Actual sizes in case of a Poi($\lambda$)-INAR(1) DGP when testing for $H_0^\text{semi}$ in \eqref{eq:null} with $p=1$.}
\begin{small}
\begin{tabular}{rr|rr|rr|rr}
\toprule
            &       & \multicolumn{2}{c|}{$a=0$}& \multicolumn{2}{c|}{$a=5$}& \multicolumn{2}{c}{$a=2$} \\
 $\lambda$ & $\alpha$ & $n=100$ & $n=500$            & $n=100$ & $n=500$            & $n=100$ & $n=500$  \\ 
  \midrule
 1 & 0.1 & 0.027 & 0.031 & 0.024 & 0.015 & 0.025 & 0.022  \\ 
  1 & 0.9 & 0.042 & 0.050 & 0.051 & 0.047 & 0.046 & 0.047  \\ 
  3 & 0.1 & 0.037 & 0.040 & 0.034 & 0.030 & 0.035 & 0.032  \\ 
  3 & 0.9 & 0.031 & 0.027 & 0.033 & 0.026 & 0.031 & 0.027 \\ 
\bottomrule
\end{tabular}
\end{small}
\label{tab:poiinar1_rev}
\end{table}

\begin{table}[h]
\centering
\caption{Power in case of a Poi($\lambda$)-INAR(2) DGP when testing for $H_0^\text{semi}$ in \eqref{eq:null} with $p=1$.}
\begin{small}
\begin{tabular}{rrr|rr|rr|rr}
\toprule
    & &       & \multicolumn{2}{c|}{$a=0$}& \multicolumn{2}{c|}{$a=5$}& \multicolumn{2}{c}{$a=2$} \\
 $\lambda$ & $\alpha_1$ & $\alpha_2$  & $n=100$ & $n=500$            & $n=100$ & $n=500$            & $n=100$ & $n=500$  \\ 
  \midrule
 1 & 0.05 & 0.05  & 0.039 & 0.034 & 0.031 & 0.020 & 0.033 & 0.022  \\ 
  1 & 0.4 & 0.5 & 0.103 & 0.311 & 0.198 & 0.657 & 0.164 & 0.538  \\ 
\bottomrule
\end{tabular}
\end{small}
\label{tab:poiinar2_rev_s1}
\end{table}

\begin{table}[h]
\centering
\caption{Actual sizes in case of a Poi($\lambda$)-INAR(1) DGP when testing for $H_0^\text{semi}$ in \eqref{eq:null} with $p=1$ using test statistic \eqref{eq:tns} with $s=2$. Numbers in italic display the power values taken from Table~1 in \citet{meikar} who test for $H_0^\text{para}$ in \eqref{eq:par_null} with $p=1$ and $G_\lambda$=Poi($\lambda$) not using a higher-order test statistic analogously to \eqref{eq:tns}.}
\begin{small}
\begin{tabular}{rr|rr|rr|rr|rr}
\toprule
            &       & \multicolumn{2}{c|}{$a=0$}& \multicolumn{2}{c|}{$a=5$}& \multicolumn{2}{c|}{$a=2$}&\multicolumn{2}{c}{\textit{MK}, $a=2$} \\
 $\lambda$ & $\alpha$ & $n=100$ & $n=500$            & $n=100$ & $n=500$            & $n=100$ & $n=500$            & $n=100$ & $n=500$ \\ 
  \midrule
  1 & 0.3 & 0.047 & 0.047 & 0.047 & 0.041 & 0.044 & 0.044 & \textit{0.049} & \textit{0.055} \\ 
   1 & 0.5 & 0.039 & 0.047 & 0.042 & 0.051 & 0.038 & 0.050 & \textit{0.051} & \textit{0.053} \\ 
   3 & 0.3 & 0.027 & 0.035 & 0.048 & 0.036 & 0.042 & 0.037 & \textit{0.056} & \textit{0.047} \\ 
   3 & 0.5 & 0.027 & 0.030 & 0.043 & 0.037 & 0.038 & 0.037 & \textit{0.055} & \textit{0.059} \\ 
  \bottomrule
\end{tabular}
\end{small}
\label{tab:poiinar1_p2}
\end{table}

\begin{table}[h]
\centering
\caption{Actual sizes in case of a NB($N,\pi$)-INAR(1) DGP when testing for $H_0^\text{semi}$ in \eqref{eq:null} with $p=1$ using test statistic \eqref{eq:tns} with $s=2$. Numbers in italic display the power values taken from Table~1 in \citet{meikar} who test for $H_0^\text{para}$ in \eqref{eq:par_null} with $p=1$ and $G_\lambda$=Poi($\lambda$) not using a higher-order test statistic analogously to \eqref{eq:tns}.}
\begin{small}
\begin{tabular}{rrr|rr|rr|rr|rr}
\toprule
     &        &       & \multicolumn{2}{c|}{$a=0$}& \multicolumn{2}{c|}{$a=5$}& \multicolumn{2}{c|}{$a=2$}&  \multicolumn{2}{c}{\textit{MK}, $a=2$} \\
 $N$ & $\pi$ & $\alpha$ & $n=100$ & $n=500$ & $n=100$ & $n=500$ & $n=100$ & $n=500$ & $n=100$ & $n=500$ \\ 
  \midrule
  1 & 1/2 & 0.5 & 0.053 & 0.053 & 0.052 & 0.051 & 0.053 & 0.052 & \textit{0.686} & \textit{1.000} \\ 
   2 & 2/3 & 0.5 & 0.046 & 0.050 & 0.048 & 0.050 & 0.049 & 0.050 &\textit{0.327 }& \textit{0.897} \\ 
   10 & 10/11 & 0.5 & 0.045 & 0.044 & 0.047 & 0.047 & 0.045 & 0.041 &\textit{0.120} & \textit{0.149} \\ 
  \bottomrule
\end{tabular}
\end{small}
\label{tab:nbinar1_p2}
\end{table}

\begin{table}[h]
\centering
\caption{Power in case of a Poi($\lambda$)-INAR(2) DGP when testing for $H_0^\text{semi}$ in \eqref{eq:null} with $p=1$ using test statistic \eqref{eq:tns} with $s=2$.}
\begin{small}
\begin{tabular}{rrr|rr|rr|rr}
\toprule
    & &          & \multicolumn{2}{c|}{$a=0$}& \multicolumn{2}{c|}{$a=5$}& \multicolumn{2}{c}{$a=2$} \\
 $\lambda$ & $\alpha_1$ & $\alpha_2$  & $n=100$ & $n=500$            & $n=100$ & $n=500$            & $n=100$ & $n=500$  \\ 
  \midrule
  1 & 0.05 & 0.05 & 0.042 & 0.087 & 0.041 & 0.114 & 0.040 & 0.111  \\ 
  1 & 0.4 & 0.5 & 0.131 & 0.515 & 0.384 & 0.983 & 0.268 & 0.900 \\ 
\bottomrule
\end{tabular}
\end{small}
\label{tab:poiinar2_rev_s2}
\end{table}

\section{Proofs}

\subsection{Proof of Lemma \ref{lemma_pgf_inarp}} \label{pr_lemma_pgf_inarp} \noindent
Exploiting the INAR($p$) model structure of $X_t, \ldots, X_{t-p}$, we get 
\begin{align*}
E\left(\prod_{j=0}^p u_j^{X_{t-j}}\right) &= E\left(E\left(\prod_{j=0}^p u_j^{X_{t-j}}|X_{t-1},\ldots,X_{t-p}\right)\right)	\\
&= E\left(\prod_{j=1}^p u_j^{X_{t-j}}E\left(u_0^{X_t}|X_{t-1},\ldots,X_{t-p}\right)\right).
\end{align*}
Taking a closer look at the interior conditional expectation and inserting the model equation, we get
\begin{align*}
 E\left(u_0^{X_t}|X_{t-1},\ldots,X_{t-p}\right) &= E\left(u_0^{\alpha_1\circ X_{t-1}+\cdots+\alpha_p\circ X_{t-p}+\varepsilon_t}|X_{t-1},\ldots,X_{t-p}\right)	\\
&= \left\{\prod_{i=1}^p E\left(u_0^{\alpha_i\circ X_{t-i}}|X_{t-1},\ldots,X_{t-p}\right)\right\}\cdot E(u_0^{\varepsilon_t})	\\
&= \left\{\prod_{i=1}^p E\left(u_0^{\alpha_i\circ X_{t-i}}|X_{t-i}\right)\right\}\cdot E(u_0^{\varepsilon_t})	\\
&= \left\{\prod_{i=1}^p \Big(1+\alpha_i(u_0-1)\Big)^{X_{t-i}}\right\}\cdot g_{\varepsilon}(u_0),
\end{align*}
where we have used that given $X_{t-1}, \ldots, X_{t-p}$, the thinning operations and the innovation $\varepsilon_t$ are independent. The thinning operation $\alpha_i \circ$ only depends on $X_{t-i}$ and $\alpha_i \circ X_{t-i} | X_{t-i} \sim \text{Bin}(X_{t-i}, \alpha_i)$ together with the well-known formula for the pgf of a binomial distribution. Furthermore, we define $g_{\varepsilon}(u_0) := E(u_0^{\varepsilon_t})$ as the marginal pgf of the innovations. Hence, we get the following representation for the joint pgf of $X_t, \ldots, X_{t-p}$:
\begin{align*}
g_{p}(u_0,\ldots,u_p) &\nonumber
:= E\left(\prod_{j=0}^p u_j^{X_{t-j}}\right)	\\
&\nonumber
= E\left(\prod_{j=1}^p u_j^{X_{t-j}}\left\{\prod_{i=1}^p (1+\alpha_i(u_0-1))^{X_{t-i}}\right\}\cdot g_{\varepsilon}(u_0)\right)	\\
&\nonumber
= g_{\varepsilon}(u_0)\cdot  E\left(\prod_{j=1}^p u_j^{X_{t-j}}\left\{\prod_{i=1}^p (1+\alpha_i(u_0-1))^{X_{t-i}}\right\}\right)	\\
&= g_{\varepsilon}(u_0)\cdot  E\left(\prod_{j=1}^p \Big\{u_j \big(1+\alpha_j(u_0-1)\big)\Big\}^{X_{t-j}}\right).
\end{align*}
\hfill $\square$

\subsection{Proof of Lemma \ref{lemma_tn_a_p}} \label{pr_lemma_tn_a_p} \noindent
Plugging-in \eqref{eq:pgfINARsp} and \eqref{eq:pgfINARp_genest} and rearranging terms, we get 
\begin{align*}
T_n =& n\int_0^1 \cdots \int_0^1 \Big(\widehat g_{p; H_0}(u_0,\ldots,u_p)-\widehat g_{p}(u_0,\ldots,u_p)\Big)^2\, w(u_0, \ldots, u_p;a) \, du_0\cdots du_p\\
=& \frac{n}{(n-p)^2} (a+1)^{p+1} \,\sum_{t,s=p+1}^{n} \int_0^1 \cdots \int_0^1 \left(\prod_{j=1}^p u_j^{X_{t-j}}\right) \left(\prod_{j=1}^p u_j^{X_{s-j}}\right)\, \left(\prod\limits_{j=1}^p u_j^a \right) \, du_1\cdots du_p\\
& \times \int_0^1 \left(u_0^{X_t}-\widehat g_{\varepsilon}(u_0)\, \prod_{j=1}^p \big(1+\widehat \alpha_{\text{sp},j}(u_0-1)\big)^{X_{t-j}}\right) \\ & \left(u_0^{X_s}-\widehat g_{\varepsilon}(u_0)\, \prod_{j=1}^p \big(1+\widehat \alpha_{\text{sp},j}(u_0-1)\big)^{X_{s-j}}\right)\, u_0^a \, du_0\\
=& \frac{n}{(n-p)^2} (a+1)^{p+1} \,\sum_{t,s=p+1}^{n} \left(\prod_{j=1}^p \int_0^1 u_j^{X_{t-j}+X_{s-j}+a}du_j\right) \\
& \times \int_0^1 \Biggl(u_0^{X_t+X_s +a} +\widehat g_{\varepsilon}(u_0)^2\, u_0^a\, \prod_{j=1}^p \big(1+\widehat \alpha_{\text{sp},j}(u_0-1)\big)^{X_{t-j}+X_{s-j}}\\
&
-\widehat g_{\varepsilon}(u_0)\, u_0^{X_s+a}\prod_{j=1}^p \big(1+\widehat \alpha_{\text{sp},j}(u_0-1)\big)^{X_{t-j}}
-\widehat g_{\varepsilon}(u_0)\, u_0^{X_t+a}\prod_{j=1}^p \big(1+\widehat \alpha_{\text{sp},j}(u_0-1)\big)^{X_{s-j}}
\Biggl)\,du_0.
\end{align*}
Furthermore, by using that
\begin{align} \label{int_rule}
\int_0^1 u^x\,du=1/(1+x)
\end{align}
holds for $x\not=-1$, $T_n$ can be simplified to get
\begin{align*}
T_n =& \frac{n}{(n-p)^2} (a+1)^{p+1} \,\sum_{t,s=p+1}^{n} \left(\prod_{j=1}^p \frac{1}{1+X_{t-j}+X_{s-j} +a}\right) \\
& \times\left(\frac{1}{1+X_{t}+X_{s} +a}
+\int_0^1 \widehat g_{\varepsilon}(u_0)^2 u_0^a\, \prod_{j=1}^p \big(1+\widehat \alpha_{\text{sp},j}(u_0-1)\big)^{X_{t-j}+X_{s-j}}du_0\right.\\
&\left. -2\int_0^1 \widehat g_{\varepsilon}(u_0)\, u_0^{X_s+a}\prod_{j=1}^p \big(1+\widehat \alpha_{\text{sp},j}(u_0-1)\big)^{X_{t-j}}du_0
\right).
\end{align*}
To be able to use the same integration rule \eqref{int_rule} also for the two remaining integrals above, we need to isolate the terms in $u_0$. On the one hand, we have
\begin{align*}
    \widehat{g}_\varepsilon(u_0)^2 = \sum\limits_{k_1,k_2=0}^{\max(X_1,\ldots,X_n)} \widehat{G}_{\text{sp}}(k_1)\widehat{G}_{\text{sp}}(k_2) u_0^{k_1+k_2}.
\end{align*}
By using twice the binomial theorem in each case, on the other hand, we get
\begin{align*}
    (1+\widehat{\alpha}_{\text{sp},j}(u_0-1))^{X_{t-j}+X_{s-j}} = \sum\limits_{i_j=0}^{X_{t-j}+X_{s-j}} \binom{X_{t-j}+X_{s-j}}{i_j} \widehat{\alpha}_{\text{sp},j}^{i_j} \sum \limits_{h_j=0}^{i_j} \binom{i_j}{h_j} u_0^{h_j}(-1)^{i_j-h_j}
\end{align*}
and
\begin{align*}
    (1+\widehat{\alpha}_{\text{sp},j}(u_0-1))^{X_{t-j}} = \sum\limits_{i_j=0}^{X_{t-j}} \binom{X_{t-j}}{i_j} \widehat{\alpha}_{\text{sp},j}^{i_j} \sum \limits_{h_j=0}^{i_j} \binom{i_j}{h_j} u_0^{h_j}(-1)^{i_j-h_j},
\end{align*}
respectively. Altogether, this leads to
\begin{align*}
  T_n &= \frac{n}{(n-p)^2} (a+1)^{p+1} \left( \prod \limits_{j=1}^p \frac{1}{1+X_{t-j}+X_{s-j}+a}\right) \Bigg[ \frac{1}{1+X_t+X_s+a}   \\
  &  +  \sum \limits_{k_1,k_2 = 0}^{\max(X_1,\ldots,X_n)} \widehat{G}_{\text{sp}}(k_1)\widehat{G}_{\text{sp}}(k_2) \sum\limits_{i_1=0}^{X_{t-1}+X_{s-1}} \sum\limits_{h_1=0}^{i_1} \ldots \sum\limits_{i_p=0}^{X_{t-p}+X_{s-p}} \sum\limits_{h_p=0}^{i_p}   \\
  &  \prod\limits_{j=1}^p \binom{X_{t-j}+X_{s-j}}{i_j} \widehat{\alpha}_{\text{sp},j}^{i_j} (-1)^{i_j-h_j} \binom{i_j}{h_j} \int\limits_0^1 u_0^{1+k_1+k_2+a+\sum_{m=1}^p h_m} du_0  -2 \sum\limits_{k=0}^{\max(X_1,\ldots,X_n)} \widehat{G}_{\text{sp}}(k)  \\
  &  \sum\limits_{i_1=0}^{X_{t-1}+X_{s-1}} \sum\limits_{h_1=0}^{i_1} \ldots \sum\limits_{i_p=0}^{X_{t-p}+X_{s-p}} \sum\limits_{h_p=0}^{i_p}  \prod\limits_{j=1}^p \binom{X_{t-j}}{i_j} \widehat{\alpha}_{\text{sp},j}^{i_j} (-1)^{i_j-h_j} \binom{i_j}{h_j}  \int\limits_0^1 u_0^{1+k+X_s+a+ \sum_{m=1}^p h_m }  du_0 \Bigg]. \notag
\end{align*}
Applying \eqref{int_rule} again, the assertion follows. \hfill $\square$

\subsection{Proof of Proposition \ref{Vstat}} \label{pr_Vstat} \noindent
The only aspect that remains to show is the degeneracy. Let $h$ be the kernel defined in \eqref{kernel_h}. Using $y_t=(x_t, \ldots, x_{t-p})\in\mathbb{N}_0^{p+1}$, under the null $H_0^{semi}(\theta_0)$ in \eqref{eq:null_theta_fixed}, we have 
\begin{align*}
E(h(y_t,Y_s;\theta_0)) &= \int\limits_0^1 \ldots \int \limits_0^1  \left(g_{0,\varepsilon}(u_0) \prod \limits_{j=1}^p (u_j(1+\alpha_{0,j}(u_0-1)))^{x_{t-j}}- \prod \limits_{j=0}^p u_j^{x_{t-j}}\right) \\
 & \quad \quad \quad \quad  \times E \left(g_{0,\varepsilon}(u_0) \prod \limits_{j=1}^p (u_j(1+\alpha_{0,j}(u_0-1)))^{X_{s-j}}- \prod \limits_{j=0}^p u_j^{X_{s-j}}\right) \\
 & \quad \quad \quad \quad w(u_0, \ldots, u_p;a) \, du_0 \ldots du_p \\
 &= \int\limits_0^1 \ldots \int \limits_0^1  \left(g_{0,\varepsilon}(u_0) \prod \limits_{j=1}^p (u_j(1+\alpha_{0,j}(u_0-1)))^{x_{t-j}}- \prod \limits_{j=0}^p u_j^{x_{t-j}}\right) \\
 & \quad \quad \quad \quad  \quad \times\Big( pgf_{H_0}(X_s, \ldots, X_{s-p}) - pgf(X_s, \ldots, X_{s-p}) \Big) \\
 & \quad \quad \quad \quad w(u_0, \ldots, u_p;a) \, du_0 \ldots du_p \\
 & = 0,
\end{align*}
because $pgf_{H_0}(X_s, \ldots, X_{s-p})=pgf(X_s, \ldots, X_{s-p})$ under $H_0^{semi}(\theta_0)$. That is, we are in the case of a degenerate kernel. \hfill $\square$

\subsection{Proof of Theorem \ref{limit_distr_prep}} \label{pr_limit_distr_prep} \noindent
Using Proposition \ref{Vstat}, we already have the degeneracy of the kernel \eqref{kernel_h}. In view of Theorem 1 in \citet{leucht2}, under $H_0^{semi}(\theta_0)$, the kernel $h$ even fulfills the stronger degeneracy condition
\begin{align} \label{strong_deg}
& E(h(y_t,Y_s;\theta_0)|Y_1, \ldots, Y_{s-1})    \nonumber\\ &= \int\limits_0^1 \ldots \int \limits_0^1  \left(g_{0,\varepsilon}(u_0) \prod \limits_{j=1}^p (u_j(1+\alpha_{0,j}(u_0-1)))^{x_{t-j}}- \prod \limits_{j=0}^p u_j^{x_{t-j}}\right) \\
&  \quad \quad  \times E \left(g_{0,\varepsilon}(u_0) \prod \limits_{j=1}^p (u_j(1+\alpha_{0,j}(u_0-1)))^{X_{s-j}}- \prod \limits_{j=0}^p u_j^{X_{s-j}}|Y_1, \ldots, Y_{s-1}\right) \notag \\
 & \quad \quad \quad \quad  w(u_0, \ldots, u_p;a) \, du_0 \ldots du_p \notag \\
 & = 0. \notag
\end{align}
By using the Markov property of an INAR process, this is the case, because
\begin{align*}
 & \quad \quad E \left(g_{0,\varepsilon}(u_0) \prod \limits_{j=1}^p (u_j(1+\alpha_{0,j}(u_0-1)))^{X_{s-j}}- \prod \limits_{j=0}^p u_j^{X_{s-j}}|Y_1, \ldots, Y_{s-1}\right) \\
 &= E \left(g_{0,\varepsilon}(u_0) \prod \limits_{j=1}^p (u_j(1+\alpha_{0,j}(u_0-1)))^{X_{s-j}}- \prod \limits_{j=0}^p u_j^{X_{s-j}}| X_{1-p}, \ldots, X_{s-1} \right) \\
 &= E \left(g_{0,\varepsilon}(u_0) \prod \limits_{j=1}^p (u_j(1+\alpha_{0,j}(u_0-1)))^{X_{s-j}}| X_{1-p}, \ldots, X_{s-1}\right) - E \left( \prod \limits_{j=0}^p u_j^{X_{s-j}}| X_{1-p}, \ldots, X_{s-1}\right) \\
 &= g_{0,\varepsilon}(u_0) \prod \limits_{j=1}^p (u_j(1+\alpha_{0,j}(u_0-1)))^{X_{s-j}} - \prod \limits_{j=1}^p u_j^{X_{s-j}} E(u_0^{X_s}| X_{1-p}, \ldots, X_{s-1}) \\
  &= g_{0,\varepsilon}(u_0) \prod \limits_{j=1}^p (u_j(1+\alpha_{0,j}(u_0-1)))^{X_{s-j}} - \prod \limits_{j=1}^p u_j^{X_{s-j}}g_{0,\varepsilon}(u_0) \prod\limits_{j=1}^p (1+\alpha_{0,j}(u_0-1))^{X_{s-j}} \\
  &= 0,
\end{align*}
where we used the null $H_0^{semi}(\theta_0)$ to get $E(u_0^{X_s}| X_{1-p}, \ldots, X_{s-1})=g_{0,\varepsilon}(u_0) \prod_{j=1}^p (1+\alpha_{0,j}(u_0-1))^{X_{s-j}}$. 

Furthermore, because the kernel $h$ is of a quadratic form, it is positive semidefinite. That is, for all $m \in \mathbb{N}$ and for all $c_1, \ldots, c_m \in \mathbb{R}, \, y_1, \ldots, y_m \in \mathbb{N}_0^{p+1}$, it holds $\sum_{t,s=1}^m c_t c_s h(y_t,y_s) \geq 0$. Moreover, because all the terms of the integrand, i.e., $g_\varepsilon(u_0), u_0, \ldots, u_p, \alpha_1, \ldots, \alpha_p$, are bounded by 0 from below and by 1 from above, we have
\begin{align*}
E(h(Y_0,Y_0;\theta_0)) &=  \int\limits_0^1 \ldots \int \limits_0^1 E \left( \left(g_{0,\varepsilon}(u_0) \prod \limits_{j=1}^p (u_j(1+\alpha_{0,j}(u_0-1)))^{X_{t-j}}- \prod \limits_{j=0}^p u_j^{X_{t-j}}\right)^2 \right) \\
 & \quad \quad \quad \quad w(u_0, \ldots, u_p;a) \, du_0 \ldots du_p \\
 & < \infty.
\end{align*}
With $(X_t)_{t \in \mathbb{Z}}$ being a strictly stationary and ergodic process \citep{duli}, we can apply Theorem 1 of \citet{leucht2}. Precisely, under the null $H_0^{semi}(\theta_0)$ and for $n \rightarrow \infty$, this leads to
\begin{align*}
T_n(\theta_0) \overset{d}{\longrightarrow} \sum\limits_{k=1}^\infty \lambda_k Z_k^2,
\end{align*}
where $(Z_k)_k$ is a sequence of independent standard normal random variables and $(\lambda_k)_k$ the sequence of nonzero eigenvalues of \eqref{lambda}
enumerated according their multiplicity with $(\Phi_k)_k$ the associated orthonormal eigenfunctions. \hfill $\square$

\subsection{Proof of Theorem \ref{limit_distr}} \label{pr_limit_distr} \noindent
Let $\theta_0 = (\boldsymbol{\alpha}_0,G_0) \in A \times \mathcal{G}$ denote the true parameter. For any $\theta \in \Theta$, we see that the kernel $h$ can be represented as 
\begin{align*}
h(x,y;\theta) = \int_{[0,1]^{p+1}} h_1(x,\mathbf{u};\theta) h_1(y,\mathbf{u};\theta) Q(d\mathbf{u}),
\end{align*}
where $\mathbf{u}=(u_0, \ldots, u_p)$, $h_1: \mathbb{R}^{p+1} \times [0,1]^{p+1} \times \Theta \rightarrow \mathbb{R}$, $\Theta = (0,1)^p \times \mathcal{G}$ with 
\begin{align} \label{h1}
h_1(y_t,\mathbf{u};\theta) = g_\varepsilon(u_0) \prod \limits_{j=1}^p (u_j(1+\alpha_j(u_0-1)))^{x_{t-j}}-\prod\limits_{j=0}^p u_j^{x_{t-j}} 
\end{align}
with $y_t=(x_t,\ldots,x_{t-p})$ and the probability measure $Q$ has probability density function $w$, that is, $dQ/d\mathbf{u}=w(\mathbf{u})$. Using that both terms of the difference in \eqref{h1} only take values in $[0,1]$, $\boldsymbol{\alpha} \in (0,1)^p$, $G \in \mathcal{G}$ and that the integration limits of all the following integrals are 0 and 1, we see that \eqref{h1} fulfills 
\[  \int\limits_{[0,1]^{p+1}} h_1(y,\mathbf{u};\theta_0)^2 Q (d\mathbf{u}) < \infty,  \quad   \int\limits_{[0,1]^{p+1}} E_{\theta_0} \left( h_1(Y_0,\mathbf{u};\theta_0)^2 \right) Q (d\mathbf{u}) < \infty  \]
as well as the continuity condition
\[  \int\limits_{[0,1]^{p+1}} \big(h_1(y,\mathbf{u};\theta_0)-h_1(\widetilde{y},\mathbf{u};\theta_0)\big)^2 Q(d\mathbf{u}) \rightarrow 0 \quad \text{for} \;\widetilde{y}-y \rightarrow 0.  \]
Due to \eqref{strong_deg}, we can conclude that
$E_{\theta_0} \left( h_1(Y_t,\mathbf{u};\theta_0) | Y_{t-1}, Y_{t-2}, \ldots \right) = 0$ holds as well.\footnote{We point out that in the original paper  of \citet{leucht2}, this condition contains a small typo.} Furthermore, the function $h_1$ in \eqref{h1} is continuously differentiable with respect to $\theta$. For the derivatives with respect to the model coefficients $\alpha_l, \, l \in \{1, \ldots, p\}$, we have
\begin{align} \label{deriv_alpha}
\frac{\partial}{\partial \alpha_l} h_1(y_t,\mathbf{u};\theta) &=  \frac{\partial}{\partial \alpha_l} g_\varepsilon(u_0) \prod \limits_{j=1}^p (u_j(1+\alpha_j(u_0-1)))^{x_{t-j}}   \\
&=  g_\varepsilon(u_0) \left(\prod \limits_{j=1, \,j \neq l}^p (u_j(1+\alpha_j(u_0-1)))^{x_{t-j}}\right) \frac{\partial}{\partial \alpha_l} (u_l(1+\alpha_l(u_0-1)))^{x_{t-l}} \notag  \\
&= g_\varepsilon(u_0) \left(\prod \limits_{j=1, \,j \neq l}^p (u_j(1+\alpha_j(u_0-1)))^{x_{t-j}}\right) x_{t-l}(u_l(1+\alpha_l(u_0-1)))^{x_{t-l}-1} u_l (u_0-1) \notag \\
&= \frac{g_\varepsilon(u_0)(u_0-1)}{1+\alpha_l(u_0-1)}  x_{t-l} \prod \limits_{j=1}^p (u_j(1+\alpha_j(u_0-1)))^{x_{t-j}}. \notag
\end{align}
Recalling that $g_\varepsilon(u_0) = \sum_{k=0}^\infty G(k) u_0^k$, for the partial derivatives with respect to the entries of the pmf of the innovation distribution $G$, that is, $(G(k), \, k \in \mathbb{N}_0)$, we get
\begin{align} \label{deriv_g}
\frac{\partial}{\partial G(k)} h_1(y_t,\mathbf{u}, \theta)&= \frac{\partial}{\partial G(k)} g_\varepsilon(u_0) \prod \limits_{j=1}^p (u_j(1+\alpha_j(u_0-1)))^{x_{t-j}}  \\
&=  u_0^k \prod \limits_{j=1}^p (u_j(1+\alpha_j(u_0-1)))^{x_{t-j}}, \notag
\end{align}  
which does not depend anymore on $(G(k)$, $k\in\mathbb{N}_0)$. With the same arguments as used for \eqref{h1}, also $\dot h_1$, the Fréchet derivative of $h_1$ with respect to $\theta$, fulfills 
\[ E_{\theta_0} \left( \int\limits_{[0,1]^{p+1}} ||\dot h_1(Y_0,\mathbf{u};\theta_0)||^2_2 Q(d\mathbf{u}) \right) < \infty, \]
\[ \int\limits_{[0,1]^{p+1}}  ||\dot h_1(y,\mathbf{u};\theta_0) - \dot h_1(\widetilde{y} ,\mathbf{u};\theta_0)||^2_2 Q(d\mathbf{u}) \rightarrow 0 \quad \text{for} \;\widetilde{y}-y \rightarrow 0. \]
Moreover, $\dot h_1$ fulfills a Lipschitz-type condition in $\theta$ which we outline in the following. For simplicity, we set $p=1$ but the subsequent arguments can be extended to higher order $p>1$. 
Denote $\widetilde\theta = (\widetilde\alpha, \widetilde G)$. Then, we get \[ \dot h_1(\cdot,\cdot,\theta)  -\dot h_1(\cdot,\cdot, \tilde\theta) = \ddot h_1(\cdot,\cdot, \check\theta)  (\theta - \tilde \theta),   \] where $\check\theta$ is between $\theta$ and $\widetilde\theta$ and
\[ \ddot h_1(\cdot,\cdot, \check\theta)(\theta-\tilde\theta) =  \frac{\partial \dot h_1(\cdot,\cdot,\theta)}{\partial \alpha}_{|\theta=\check\theta} \, (\alpha-\tilde\alpha) +  \sum\limits_{k=0}^\infty \frac{\partial \dot h_1(\cdot,\cdot,\theta)}{\partial G(k)}_{|\theta=\check\theta} \, (G(k)-\tilde G(k)) 
\] with 
\begin{align} \label{eq:doth1_alpha}
\frac{\partial \dot h_1(Y_t,\mathbf{u},\theta)}{\partial \alpha}_{|\theta=\check\theta} =  \begin{pmatrix}
\check g_\varepsilon(u_0) (u_0-1)^2 X_{t-1}u_1^{X_{t-1}} (X_{t-1}-1) (1+\check\alpha(u_0-1))^{X_{t-1}-2}\\
u_0^0 u_1^{X_{t-1}} X_{t-1} (1+\check\alpha(u_0-1))^{X_{t-1}-1}(u_0-1) \\
u_0^1 u_1^{X_{t-1}} X_{t-1} (1+\check\alpha(u_0-1))^{X_{t-1}-1}(u_0-1)  \\
\vdots 
\end{pmatrix} \end{align}
and
\begin{align} \label{eq:doth1_G}
\frac{\partial \dot h_1(Y_t,\mathbf{u},\theta)}{\partial G(k)}_{|\theta=\check\theta} =  \begin{pmatrix}
u_0^k(u_0-1)X_{t-1}u_1^{X_{t-1}}(1+\check\alpha(u_0-1))^{X_{t-1}-1}\\
0 \\
0 \\
\vdots
    \end{pmatrix},\quad k \in \mathbb{N}_0, \end{align} 
where we used the previous calculations of \eqref{deriv_alpha} and \eqref{deriv_g}.
For the first entry of \eqref{eq:doth1_alpha}, we have 
\[ \check g_\varepsilon(u_0) (u_0-1)^2 X_{t-1}u_1^{X_{t-1}} (X_{t-1}-1) (1+\check\alpha(u_0-1))^{X_{t-1}-2} \leq X_{t-1}(X_{t-1}-1) \leq X_{t-1}^2 \] and for all other entries (which are equal to the first entry of \eqref{eq:doth1_G}, that is, for all $j\in\mathbb{N}$, we have \[ u_0^j(u_0-1)X_{t-1}u_1^{X_{t-1}}(1+\check\alpha(u_0-1))^{X_{t-1}-1} \leq X_{t-1} \leq X_{t-1}^2. \]
Hence, using the notation $\|(a_n)_{n\in\mathbb{N}}\|_1=\sum_{n=1}^\infty |a_n|$, we get
\begin{align*}
|| \dot h_1(Y_t,\mathbf{u},\theta)  -\dot h_1(Y_t,\mathbf{u}, \widetilde\theta) ||_1 &= ||\ddot h_1(Y_t,\mathbf{u}, \widetilde\theta)  (\theta - \widetilde \theta) ||_1 \\
& \leq \left \lVert \begin{pmatrix}
    X_{t-1}^2 (\alpha - \widetilde\alpha) + \sum\limits_{k=0}^\infty  X_{t-1}^2 (G(k)-\widetilde G(k)) \\
     X_{t-1}^2 (G(0)-\widetilde G(0)) \\
     X_{t-1}^2 (G(1)-\widetilde G(1)) \\
     \vdots
\end{pmatrix} \right \rVert_1 \\
& \leq X_{t-1}^2 \left \lVert \begin{pmatrix}
    ||\theta-\widetilde \theta||_1 \\
     ||G-\widetilde G||_1
\end{pmatrix} \right \rVert_1 \\
& \leq 2 X_{t-1}^2  ||\theta- \widetilde \theta||_1,
\end{align*}
as $||G-\widetilde G||_1\leq||\theta-\widetilde \theta||_1$. That is, the last part of Assumption (A2) (iv) in \citet{leucht2} holds for existing second moments of $X_t$ ensured by Assumption \ref{gtildeg}. 
Additionally, \citet{drost} show that their proposed estimator \eqref{eq:sp_estimator} is regular and consistent and consequently exhibits the Bahadur-type expansion
\begin{align*}
\widehat{\theta}_\text{sp} = \theta_0 + \frac{1}{n} \sum\limits_{t=1}^n l_t + o_p(n^{-1/2}),
\end{align*}
with $l_t=L(Y_t, Y_{t-1}, \ldots)$ for some measurable function $L$, $E_{\theta_0}(l_t|Y_{t-1},Y_{t-2}, \ldots) = 0$ and $E_{\theta_0}(||l_t||_2^2) < \infty$, where we refer to Section 5.3 of \citet{vandervaart} proving this result for the class of M-estimators (including ML-estimators) under mild regularity conditions. Hence, all assumptions of Proposition 1 in \citet{leucht2}, which is based on \citet{dewet}, are fulfilled and we get that $T_n(\widehat{\theta}_{\text{sp}})$ has the same limiting distribution as $\frac{1}{n} \sum_{t=1}^n \sum_{s=1}^n \widetilde{h}(\widetilde{Y}_t,\widetilde{Y}_s;\theta_0)$, where $\widetilde{Y}_t=(Y_t',l_t')'$ and 
\begin{align} \label{kernel_hhat}
\widetilde{h}(x,y;\theta_0)= \int\limits_{[0,1]^{p+1}} \hspace*{-0.2cm}\big(h_1(x_1,\mathbf{u};\theta_0) + E_{\theta_0}(\dot h_1(Y_1,\mathbf{u};\theta_0) x_2)) (h_1(y_1,\mathbf{u};\theta_0) + E_{\theta_0}(\dot h_1(Y_1,\mathbf{u};\theta_0) y_2)\big) Q(d\mathbf{u}).
\end{align}
Altogether, using Theorem 1 of \citet{leucht2}, we get 
\begin{align*}
T_n =  T_n(\widehat{\theta}_\text{sp}) \overset{d}{\longrightarrow}  \sum\limits_{k=1}^\infty \widetilde{\lambda}_k Z_k^2,
\end{align*}
where $(Z_k)_k$ is as before and $(\widetilde{\lambda}_k)_k$ denotes the sequence of nonzero eigenvalues of the equation \eqref{lambda_tilde}
enumerated according their multiplicity with $(\widetilde{\Phi}_k)_k$ the associated orthonormal eigenfunctions. 
\hfill $\square$

\subsection{Proof of Theorem \ref{test_cons}} \label{pr_test_cons} \noindent
Let $p\in\mathbb{N}$ and suppose we observe data $X_1,\ldots,X_n$ from some (strictly) stationary count time series process $(X_t,t\in\mathbb{Z})$ under the alternative $H_1^{semi}$ in  \eqref{eq:alternative}
such that \eqref{convergence_misspecification}, \eqref{eq:fixed_alternative_pgf} and \eqref{weak_convergence_fixed_alternative} hold. 
Then, by adding suitable zeros $g_{p;H_0}(\mathbf{u})-g_{p;H_0}(\mathbf{u})$ and $g_{p}(\mathbf{u})-g_{p}(\mathbf{u})$, where $\mathbf{u}=(u_0,\ldots,u_p)$, to the integrand of the test statistic $T_n$ from \eqref{eq:tn_int} and expanding the squared term in brackets, we get
\begin{align*} 
T_n =& n\int_0^1 \cdots \int_0^1 \Big(\widehat g_{p; H_0}(\mathbf{u})-\widehat g_{p}(\mathbf{u})\Big)^2 \, w(\mathbf{u};a) \, du_0\cdots du_p \notag \\
=& n\int_0^1 \cdots \int_0^1 \Big(\widehat g_{p; H_0}(\mathbf{u})+\big(g_{p; H_0}(\mathbf{u})-g_{p; H_0}(\mathbf{u})\big)-\widehat g_{p}(\mathbf{u})+\big(g_{p}(\mathbf{u})-g_{p}(\mathbf{u})\big)\Big)^2 \notag \\ & \qquad \qquad \qquad w(\mathbf{u};a) \, du_0\cdots du_p  \notag \\
=& n\int_0^1 \cdots \int_0^1 \Big(\widehat g_{p; H_0}(\mathbf{u})-g_{p; H_0}(\mathbf{u})-\widehat g_{p}(\mathbf{u})+ g_{p}(\mathbf{u})\Big)^2 w(\mathbf{u};a) \, du_0\cdots du_p    \\
&+ n\int_0^1 \cdots \int_0^1 \Big(g_{p; H_0}(\mathbf{u})- g_{p}(\mathbf{u})\Big)^2\, w(\mathbf{u};a) \, du_0\cdots du_p  \notag \\
&+ n\int_0^1 \cdots \int_0^1 \Big(\widehat g_{p; H_0}(\mathbf{u})-g_{p; H_0}(\mathbf{u})-\widehat g_{p}(\mathbf{u})+ g_{p}(\mathbf{u})\Big)\Big(g_{p; H_0}(\mathbf{u})- g_{p}(\mathbf{u})\Big)\, w(\mathbf{u};a) \, du_0\cdots du_p  \notag   \\
=:&\,  A_1 +A_2 +A_3   \notag
\end{align*}
with an obvious notation for $A_i$, $i=1,2,3$. When discussing these three terms separately, we see that the integrand of the first term $A_1$ is appropriately centered such that $A_1$ represents again a degenerate V-statistic (just with a different kernel) that converges to a (non-degenerate) $\chi^2$-type distribution by making use of \eqref{weak_convergence_fixed_alternative}. The second term $A_2$ diverges to $+\infty$ as, for all $\mathbf{u}\in[0,1]^{p+1}$, $g_{p; H_0}(\mathbf{u})- g_{p}(\mathbf{u})$ converges to a non-zero limit such that its square becomes a function that will be strictly positive on some subset of $[0,1]^{p+1}$ with strictly positive Lebesgue measure according to \eqref{eq:fixed_alternative_pgf}. 
This leads to an integral that is also strictly positive and as it is inflated with $n$, this second term diverges to $+\infty$ with rate $n$. Finally, the third term is a mixed term, where the difference in the first brackets, that is, $\widehat g_{p; H_0}(\mathbf{u})-g_{p; H_0}\mathbf{u})-\widehat g_{p}(\mathbf{u})+ g_{p}(\mathbf{u})$, is of order $O(1/\sqrt{n})$. This is multiplied with $g_{p; H_0}(\mathbf{u})- g_{p}(\mathbf{u})$, which is a function that is strictly positive on some set with positive Lebesgue measure. In total, the integral behaves like $O_P(1/\sqrt{n})$ such that the whole third term, when inflated with $n$ behaves like $O_P(\sqrt{n})$. As this is slower than the rate $O(n)$ obtained for the second term, altogether, we have $T_n=O_P(n)$. That is, $T_n$ diverges to $+\infty$ in probability such that, for all $\gamma \in(0,1)$, we have $E(\varphi_n)=P(T_n>q_{1-\gamma})\longrightarrow 1$. Hence, this proves consistency of the test against fixed alternatives.
$\hfill \square$

\subsection{Proof of Theorem \ref{test_local_power}} \label{pr_test_local_power} \noindent
Let $p\in\mathbb{N}$ and suppose we observe data $X_{n,1},\ldots,X_{n,n}$ from a triangular array $(X_{n,t}, t=1,\ldots,n,\ n\in\mathbb{N})$ of count time series and, for each fixed $n$, $X_{n,1},\ldots,X_{n,n}$ is generated from a stationary Markov chain of order $p$ with state space $\mathcal{S}\subseteq \mathbb{N}_0$ generated under the alternative $H_1^{\text{semi}}$ such that \eqref{eq:local_alternative_pgf} with $a_n=n^{-1/2}$ as well as \eqref{weak_convergence} hold. Then, similar to the proof of Theorem \ref{test_cons}, by adding suitable zeros $g_{p;H_0,n}(\mathbf{u})-g_{p;H_0,n}(\mathbf{u})$ and $g_{p,n}(\mathbf{u})-g_{p,n}(\mathbf{u})$
to the integrand of the test statistic $T_n$ from \eqref{eq:tn_int} and expanding the squared term in brackets, we get
\begin{align} \label{eq:tn_zero_add_local}
T_n =& n\int_0^1 \cdots \int_0^1 \Big(\widehat g_{p; H_0,n}(\mathbf{u})-\widehat g_{p,n}(\mathbf{u})\Big)^2 \, w(\mathbf{u};a) \, du_0\cdots du_p \notag \\
=& n\int_0^1 \cdots \int_0^1 \Big(\widehat g_{p; H_0,n}(\mathbf{u})+\big(g_{p; H_0,n}(\mathbf{u})-g_{p; H_0,n}(\mathbf{u})\big)-\widehat g_{p,n}(\mathbf{u})+\big(g_{p,n}(\mathbf{u})-g_{p,n}(\mathbf{u})\big)\Big)^2 \notag \\
& \qquad\qquad \qquad w(\mathbf{u};a) \, du_0\cdots du_p  \notag \\
=& \int_0^1 \cdots \int_0^1 \Big(\sqrt{n}\big(\widehat g_{p;H_0,n}(\mathbf{u})-\widehat g_{p,n}(\mathbf{u})-\left(g_{p;H_0,n}(\mathbf{u})-g_{p,n}(\mathbf{u})\right)\big)+\sqrt{n}\big(g_{p; H_0,n}(\mathbf{u})-g_{p,n}(\mathbf{u})\big)\Big)^2 \notag \\
& \qquad\qquad \qquad w(\mathbf{u};a) \, du_0\cdots du_p  \notag \\
\overset{d}{\longrightarrow}& \int_0^1 \cdots \int_0^1 \Big(G(\mathbf{u})+C(\mathbf{u})\Big)^2 w(\mathbf{u};a) \, du_0\cdots du_p,  \notag
\end{align}
where we used \eqref{eq:local_alternative_pgf} with $a_n=1/\sqrt{n}$ and \eqref{weak_convergence}. Otherwise, if $a_n\rightarrow 0$ such that $\sqrt{n} \, a_n\rightarrow \infty$, the test $\varphi_n$ remains consistent, that is, we have $E(\varphi_n)\rightarrow 1$ as $n\rightarrow \infty$. If $a_n=o(n^{-1/2})$, the test $\varphi_n$ has no asymptotic power, that is, we have $E(\varphi_n)\rightarrow \gamma$ as $n\rightarrow \infty$.

$\hfill \square$

\end{appendix}

\end{document}